\documentclass[twocolumn]{aastex631}

\usepackage{tabularx}
\usepackage{url}
\usepackage{graphicx}
\usepackage[T1]{fontenc}
\usepackage{ae,aecompl}
\usepackage{booktabs}
\usepackage{natbib}
\usepackage{multirow}
\usepackage{multirow}
\usepackage{appendix}
\usepackage{blindtext}
\usepackage{longtable}
\usepackage{tabu}
\usepackage{lineno}

\usepackage{chngcntr}

\def\unit #1{\,{\rm #1}}

\newcommand\kms{\rm \,\unit{km\,s^{-1}}}

\newcommand\kev{\rm \,\unit{keV}}

\newcommand\erg{\rm \,\unit{erg\,}}

\newcommand\funit{\rm \,erg\,cm^{-2}\,s^{-1}}

\newcommand\lunit{\rm \,erg \,s^{-1}}

\newcommand\ledd{L_{\rm \, Edd}}

\newcommand\lambdaedd{\lambda_{\rm \, Edd}}

\newcommand\lbol{L_{\rm \, bol}}

\newcommand\funita{\rm\,erg\,cm^{-2}\,s^{-1}\angstrom^{-1}}

\newcommand\msol{M_{\odot}}

\newcommand\msolyi{M_{\odot}\,\rm yr^{-1}}

\newcommand\pc{\unit{pc}}

\newcommand\mpc{\unit{Mpc}}

\newcommand\ev{\unit{\, eV}}

\newcommand{\angstrom}{\mbox{\normalfont\AA}}

\newcommand\swift{{\it Swift}}
\newcommand\xmm{{\it XMM-Newton}}

\newcommand\nicer{{\it NICER}}

\newcommand\cl{{CL-AGN}}
\newcommand{\angs}{\mbox{\normalfont\AA}}

\def\aap{A\&A} \def\apjl{ApJ}  
 \def\apjs{ApJS}

 \submitjournal{ApJ}

\shorttitle{Changing look AGN 1ES 1927+654}

\begin{document}

\title{A re-emerging bright soft-X-ray state of the changing-look Active Galactic Nucleus 1ES~1927+654: a multi-wavelength view}

\author[0000-0003-4790-2653]{Ritesh Ghosh}
\affiliation{Center for Space Science and Technology, University of Maryland Baltimore County, 1000 Hilltop Circle, Baltimore, MD 21250, USA.}
\affiliation{Astrophysics Science Division, NASA Goddard Space Flight Center, Greenbelt, MD 20771, USA.}
\affiliation{Center for Research and Exploration in Space Science and Technology, NASA/GSFC, Greenbelt, Maryland 20771, USA}

\author[0000-0003-2714-0487]{Sibasish Laha} 

\affiliation{Center for Space Science and Technology, University of Maryland Baltimore County, 1000 Hilltop Circle, Baltimore, MD 21250, USA.}
\affiliation{Astrophysics Science Division, NASA Goddard Space Flight Center, Greenbelt, MD 20771, USA.}
\affiliation{Center for Research and Exploration in Space Science and Technology, NASA/GSFC, Greenbelt, Maryland 20771, USA}

\author[0000-0000-0000-0000]{Eileen Meyer}
\affiliation{Department of Physics, University of Maryland, Baltimore County, 1000 Hilltop Circle, Baltimore, MD 21250, USA.}

\author[0000-0003-1101-8436]{Agniva Roychowdhury}
\affiliation{Department of Physics, University of Maryland, Baltimore County, 1000 Hilltop Circle, Baltimore, MD 21250, USA.}

\author[0000-0002-4439-5580]{Xiaolong Yang}
\affiliation{Shanghai Astronomical Observatory, Chinese Academy of Sciences, Shanghai 200030, China}
\affiliation{Kavli Institute for Astronomy and Astrophysics, Peking University, Beijing 100871, China}

\author[0000-0000-0000-0000]{J.~A.~Acosta--Pulido}
\affiliation{Instituto de Astrof\'isica de Canarias (IAC), E-38200 La Laguna, Tenerife, Spain}
\affiliation{Universidad de La Laguna (ULL), Departamento de Astrof\'isica, E-38206 La Laguna, Tenerife, Spain}

\author[0000-0002-8377-9667]{Suvendu Rakshit}
\affiliation{Aryabhatta  Research Institute of Observational Sciences (ARIES), Manora Peak, Nainital, 263002 India}

\author[0000-0000-0000-0000]{Shivangi Pandey}
\affiliation{Aryabhatta  Research Institute of Observational Sciences (ARIES), Manora Peak, Nainital, 263002 India}

\author[0000-0000-0000-0000]{Josefa Becerra  Gonz\'alez} 
\affiliation{Instituto de Astrof\'isica de Canarias (IAC), E-38200 La Laguna, Tenerife, Spain}
\affiliation{Universidad de La Laguna (ULL), Departamento de Astrof\'isica, E-38206 La Laguna, Tenerife, Spain}

\author[0000-0000-0000-0000]{Ehud Behar}
\affiliation{Department of Physics, Technion, Haifa 32000, Israel}

\author[0000-0000-0000-0000]{Luigi C. Gallo}
\affiliation{Department of Astronomy \& Physics, Saint Mary’s University, 923 Robie Street, Halifax, Nova Scotia, B3H 3C3, Canada}


\author[0000-0003-0543-3617]{Francesca Panessa}
\affiliation{INAF - Istituto di Astrofisica e Planetologia Spaziali, via Fosso del Cavaliere 100, I-00133 Roma, Italy}

\author[0000-0002-4622-4240]{Stefano Bianchi}
\affiliation{Dipartimento di Matematica e Fisica, Universit\`a degli Studi Roma Tre, Via della Vasca Navale 84, I-00146, Roma, Italy}

\author[0000-0002-1239-2721]{Fabio La Franca}
\affiliation{Dipartimento di Matematica e Fisica, Universit\`a degli Studi Roma Tre, Via della Vasca Navale 84, I-00146, Roma, Italy}

\author[0000-0002-1239-2721]{Nicolas Scepi}
\affiliation{School of Physics and Astronomy, University of Southampton, Highfield, Southampton, SO17 1BJ, UK.}

\author[0000-0003-0936-8488]{Mitchell C.~Begelman}
\affiliation{JILA, University of Colorado and National Institute of Standards and Technology, 440 UCB, Boulder, CO 80309-0440, USA.}


\author[0000-0002-1239-2721]{Anna Lia Longinotti}
\affiliation{Instituto de Astronomía, Universidad Nacional Autónoma de México, Circuito Exterior, Ciudad Universitaria, Ciudad de México 04510, México}

\author[0000-0003-0083-1157]{Elisabeta Lusso}
\affiliation{Dipartimento di Fisica e Astronomia, Università di Firenze, Via G. Sansone 1, 50019, Sesto Fiorentino, Firenze, Italy}
\affiliation{INAF - Osservatorio Astrofisico di Arcetri, Largo Enrico Fermi 5, 50125, Firenze, Italy}

\author[0000-0000-0000-0000]{Samantha Oates}

\affiliation{Birmingham Institute for Gravitational Wave Astronomy and School of Physics and Astronomy, University of Birmingham, Birmingham B15 2TT, UK}

\author[0000-0002-2555-3192]{Matt Nicholl}

\affiliation{Astrophysics Research Centre, School of Mathematics and Physics, Queens University Belfast, Belfast BT7 1NN, UK}

\author[0000-0003-1673-970X]{S. Bradley Cenko}
\affiliation{Astrophysics Science Division, NASA Goddard Space Flight Center, Greenbelt, MD 20771, USA.}
\affiliation{Joint Space-Science Institute, University of Maryland, College Park, MD 20742, USA}

\author[0000-0002-9700-0036]{Brendan O'Connor}
    \affiliation{Department of Physics, The George Washington University, Washington, DC 20052, USA}
    \affiliation{Astronomy, Physics and Statistics Institute of Sciences (APSIS), The George Washington University, Washington, DC 20052, USA}
    \affiliation{Department of Astronomy, University of Maryland, College Park, MD 20742-4111, USA}
    \affiliation{Astrophysics Science Division, NASA Goddard Space Flight Center, Greenbelt, MD 20771, USA.}

\author[0000-0002-5698-8703]{Erica Hammerstein}
\affiliation{Department of Astronomy, University of Maryland, College Park, MD 20742-4111, USA}
\affiliation{Astrophysics Science Division, NASA Goddard Space Flight Center, Greenbelt, MD 20771, USA.}
\affiliation{Center for Research and Exploration in Space Science and Technology, NASA/GSFC, Greenbelt, Maryland 20771, USA}

\author[0000-0000-0000-0000]{Jincen Jose}
\affiliation{Aryabhatta  Research Institute of Observational Sciences (ARIES), Manora Peak, Nainital, 263002 India}

\author[0000-0003-1020-1597]{Krisztina \'Eva Gab\'anyi}
\affiliation{Department of Astronomy, Institute of Physics and Astronomy, ELTE E\"otv\"os Lor\'and University, P\'azm\'any P\'eter s\'et\'any 1/A, H-1117 Budapest, Hungary}
\affiliation{Konkoly Observatory, ELKH Research Centre for Astronomy and Earth Sciences, Konkoly Thege Mikl\'os \'ut 15-17, H-1121 Budapest, Hungary}
\affiliation{ELKH-ELTE Extragalactic Astrophysics Research Group, ELTE E\"otv\"os Lor\'and University, P\'azm\'any P\'eter s\'et\'any 1/A, H-1117 Budapest, Hungary}

\author[0000-0001-5742-5980]{Federica Ricci}
\affiliation{Dipartimento di Matematica e Fisica, Universit\`a degli Studi Roma Tre, Via della Vasca Navale 84, I-00146, Roma, Italy}
\affiliation{INAF - Osservatorio Astronomico di Roma, via Frascati 33, 00040 Monteporzio Catone, Italy}

\author[0000-0003-1601-8048]{Sabyasachi Chattopadhyay}
\affiliation{South African Astronomical Observatory, 1 Observatory Rd, Observatory, Cape Town, 7925, South Africa}


\correspondingauthor{Ritesh Ghosh}
\email{ritesh.ghosh@nasa.gov,ritesh.ghosh1987@gmail.com}

\begin{abstract}

1ES1927+654 is a nearby active galactic nucleus that has shown an enigmatic outburst in optical/UV followed by X-rays, exhibiting strange variability patterns at timescales of months-years. Here we report the unusual X-ray, UV, and radio variability of the source in its post-flare state (Jan 2022- May 2023). Firstly, we detect an increase in the soft X-ray ($0.3-2\kev$) flux from May 2022- May 2023 by almost a factor of five, which we call the bright-soft-state. The hard X-ray $2-10\kev$ flux increased by a factor of two, while the UV flux density did not show any significant changes ($\le 30\%$) in the same period. The integrated energy pumped into the soft and hard X-ray during this period of eleven months is $\sim 3.57\times 10^{50}$ erg and $5.9\times 10^{49}$ erg, respectively. From the energetics, it is evident that whatever is producing the soft excess (SE) is pumping out more energy than either the UV or hard X-ray source. Since the energy source presumably is ultimately the accretion of matter onto the SMBH, the SE emitting region must be receiving the majority of this energy. In addition, the source does not follow  the typical disc-corona relation found in AGNs, neither in the initial flare (in 2017-2019) nor in the current bright soft state (2022-2023). We found that the core ($<1\pc$) radio emission at 5 GHz gradually increased till March 2022 but showed a dip in August 2022. The G\"udel Benz relation ($L_{\rm radio}/L_{\rm X-ray}\sim 10^{-5}$), however, is still within the expected range for radio-quiet AGN and further follow-up radio observations are currently being undertaken.

\end{abstract}


\vspace{0.5cm}


\section{Introduction}

Active Galactic Nuclei (AGN) are astrophysical systems hosting an accreting super-massive black hole (SMBH) and emitting across all wavelength bands through various physical processes. However, the exact physics of how matter from larger distances ($>1 \pc$) loses angular momentum and falls into the accretion disk of the SMBH, thereby creating enormous luminosity, is still not properly understood. Similarly, it is also not clear how the hot (T$\sim 10^9$ K) X-ray emitting plasma (corona), very commonly found in AGN, is energetically supported by the accreting system. Ideally, one would understand the physics behind these emission features when they switch on and off or show extreme variability. However, our understanding of these systems is much hindered by their long duty cycle \citep[$\sim 10^{7}-10^{9}$ years, see, e.g.,][]{Marconi2004,2015MNRAS.451.2517S} compared to the human timescale, thus preventing us from detecting an ignition or quenching event. 

However, recent large-scale time-domain surveys, e.g., All-Sky Automated Survey for Supernovae (ASAS-SN) \citep{ASAS-SN}, Zwicky Transient Facility (ZTF) \cite{ztf_survey} and others, have led to the identification of new types of extreme variability in active galaxies, called ``changing look" or ``changing state" active galactic nuclei (CL-AGNs, hereafter). These extreme variations are characterized by order-of-magnitude changes in the optical, UV, and X-ray luminosity of the source coupled with a rapid transition between spectral states (type-1 to type-2 and vice-versa). For example, sources initially exhibiting AGN type-2 characteristics with the optical/UV band dominated by narrow emission lines only have transitioned to a type-1 state, displaying prominent broad emission lines, and vice versa. There is often also a change in the optical continuum slope from red to blue coinciding with the appearance/disappearance of broad lines \citep{Lamassa2015, ruan2019}. We note here that there are two types of CL-AGNs, changing-obscuration AGNs, and changing-accretion state AGNs. The CL-AGNs we talk about in this work relate to the sources exhibiting rapid accretion state changes and not obscuration changes \citep{ricci2023}.

1ES~1927+654 (alias AT2018zf, ASASSN-18el) is one such CL-AGN that has undergone some dramatic changes \citep{trak19, Ricci2020, Ricci2021,laha2022, Li_broad_line_1es, masterson2022} in the recent past showing behaviour, unlike other known CL-AGNs. Previous optical and X-ray studies \citep{Boller2003,Gallo2013} have shown that despite having only narrow emission lines in the optical spectrum, there is no evidence of the line of sight absorption in any wavelength, thus exhibiting a `true type 2' behaviour, meaning that the BLR is simply intrinsically absent or too weak to be detected for this source. In Dec 2017, it showed a sudden rise in the optical/UV flux by almost four magnitudes which peaked around $\sim 200$ days after the start of the burst. The other interesting events that followed in the next few months-years are; broad emission lines appeared in this `true type-2 AGN' after $\sim 200$ days of the initial flare, which stayed on for another $\sim 300$ days and then gradually vanished; the X-rays gradually dimmed after $\sim 3-4$ months of the optical flare and the $2-10\kev$ hard X-rays completely vanished in Aug-2018 for almost $3$ months (Aug-2018- Oct-2018); the hard X-rays revived after Oct-2018 and flared up $\sim 10$ times its pre-flare value and stayed there for about a year; there was no correlation between the soft and the hard X-ray variability, as well as between the X-rays and the UV during this flaring event. This switching off the corona where the $2-10\kev$ hard X-ray flux was completely gone \citep{Ricci2020, Ricci2021} is particularly a peculiar phenomenon that was first witnessed in this source.
In the meantime, after the initial flare, the UV flux monotonically dimmed with $t^{-0.91\pm 0.04}$ \citep{laha2022} and returned to a near-pre-flare value after $\sim 1200$ days of the initial flare. The core ($<1\pc$) radio flux density at $5$ GHz showed a minimum (a factor of 4 below the pre-flare value) at the time when the X-ray flux was low, and it gradually increased over the next 2 years. Some studies of this phenomenon claimed it to be a tidal disruption event (TDE) \citep{Ricci2020,Ricci2021,masterson2022}. However, there were already some concerns about this description, such as the optical emission lines one would expect for a TDE like event were not present \citep{trak19}, and the X-rays did not vary in the usual way a TDE should do, that follows the UV flare with a time lag \citep{Ricci2020}. We note that some TDEs emit X-rays but do not show the same UV/X-ray evolution \citep[see, e.g., ASASSN-15oi and AT2019azh;][]{Gezari_tde_2021}. However, the vanishing and reemergence of the coronal emission of 1ES~1927+654 are uncommon in TDEs in addition to a flatter ($t^{-0.91\pm 0.04}$) UV light-curve. This result is also supported by high-cadence data from the Transiting Exoplanet Survey Satellite (TESS) survey of the source \citep{Tess_hinkle_1es1927}. The completely independent variations of the UV and the X-rays pointed to two or more different physical processes (possibly involving magnetic field in the accretion disk) apart from the accretion, which contributed to the X-ray variations. 

Our earlier study \citep{laha2022} encompassing the entire event (Dec 2017- Dec 2021) suggested that this could be a case of magnetic pole inversion of the SMBH. In such a case, matter exhibiting opposite magnetic polarity is advected to the accretion disk, which cancels the existing magnetic field, therefore temporarily switching off the corona, which is dependent on the magnetic field. Further advection of matter with the opposite polarity builds up a strong magnetic field (now with reversed poles), which re-creates the corona. Finally, the accretion rate gets back to the pre-flare value, and so does the X-ray emission \citep{scepi21,laha2022}.

Other CL-AGNs such as Mrk~590 \citep{Denney2014, Mathur2018, Ghosh_mrk590}, Mrk~1080 \citep{Mrk1018_mcElroy,noda_mrk1018}, NGC~1566 \citep{NGC1566_parker,tripathi_ngc1566} have also shown different patterns of optical, UV and X-ray variability and different timescales, mostly over several years. However, as it stands, the changing look phenomenon in 1ES~1927+654 is one of the fastest detected so far, with changes happening within just a few months. The study of these interesting sources not only helps us understand how the accretion disk and the corona get coupled (strong correlation between UV and X-rays) and decoupled (no correlation) at different phases of evolution but also how the soft X-ray emission in AGN (popularly known as the soft X-ray excess) evolves with the other components of the central engine, which is an ideal technique to understand its origin in the first place. Thus these sources give us a unique view of how the soft, hard X-ray and UV evolve with time, the origin of the soft X-ray excess, and the relation between the emission in different energy bands as they vary (e.g., radio and X-rays).

We have followed up 1ES~1927+654 with multi-wavelength observations from space-based and ground-based missions to track its behaviour in the post-flare state. In this paper, we report another interesting phenomenon currently ongoing in the source, using multi-wavelength data from \swift{} (X-ray and UV), Himalayan Chandra Telescope or HCT (optical), Lowell Discovery telescope or LDT (Optical), and Very Long Baseline Array or VLBA (radio). The paper is arranged as follows: Section \ref{sec:obs} discusses the observation, data reduction, and analysis. Section \ref{sec:results} lists the most important results. Sec \ref{sec:discussion} discusses the results and Sec \ref{sec:conclusions} lists the main conclusions. Throughout this paper, we assumed a cosmology with $H_{0} = 71\kms \mpc^{-1}, \Omega_{\Lambda} = 0.73$ and $\Omega_{M} = 0.27$. 
We regard a correlation to be significant if the confidence level is $>99.99\%$.


\begin{table*}

\centering
  \caption{The details of multi-wavelength observations of 1ES~1927+654 used in this work. Refer to \citep{laha2022} for all the previous \swift{} observations. The \xmm{} observation is used as a comparison for the pre-flare X-ray state. }\label{Table:obs}
  \begin{tabular}{cccccccc} \hline\hline 
Observation band	& Telescopes			&observation date	&observation ID	& Net exposure & Short-id	\\
			        &   				    & YYYY-MM-DD               &		        &(Sec)	\\ \hline 
 X-ray and UV     & XMM-Newton EPIC-pn/OM & 2011-05-20  &0671860201  & 28649  & X1 \\\hline
''	&{\it Swift-XRT/UVOT}&2018-05-17	& 00010682001	&2190   &S01	\\\hline
''		            & ''                 &2022-03-22		& 00010682033	&2056   &S33	\\
''		            & ''                 &2022-04-23		& 00010682034	&1276   &S34	\\
''		            & ''                 &2022-05-20		& 00010682035	&563    &S35	\\
''		            & ''                 &2022-05-26		& 00010682036	&884    &S36	\\
''		            & ''                 &2022-06-20		& 00010682037	&1469   &S37	\\
''		            & ''                 &2022-07-20		& 00010682038	&1412   &S38	\\
''		            & ''                 &2022-08-20		& 00010682039	&652    &S39	\\
''		            & ''                 &2022-08-24		& 00010682040	&1234   &S40	\\
''		            & ''                 &2022-08-26		& 00010682041	&1810   &S41	\\
''		            & ''                 &2022-09-26		& 00010682042	&2166   &S42	\\
''		            & ''                 &2022-10-28		& 00010682043	&411    &S43	\\
''		            & ''                 &2022-11-20		& 00010682044	&797    &S44	\\
''		            & ''                 &2022-11-26		& 00010682045	&1768   &S45	\\
''		            & ''                 &2022-12-03		& 00010682046	&988    &S46	\\
''		            & ''                 &2022-12-14		& 00010682047	&3001   &S47	\\
''		            & ''                 &2022-12-17		& 00010682048	&1441    &S48	\\
''		            & ''                 &2022-12-21		& 00010682049	&1989    &S49 	\\
''		            & ''                 &2023-01-03		& 00010682050	&712    &S50 	\\
''		            & ''                 &2023-01-07		& 00010682051	&860    &S51 	\\
''		            & ''                 &2023-01-14		& 00010682052	&2958   &S52 	\\
''		            & ''                 &2023-01-21		& 00010682053	&817    &S53 	\\
''		            & ''                 &2023-01-25		& 00010682054	&1907    &S54 	\\
''		            & ''                 &2023-01-28		& 00010682055   &1926    &S55 	\\
''		            & ''                 &2023-02-01    	& 00010682056	&1712    &S56 	\\
''		            & ''                 &2023-02-04		& 00010682057   &922    &S57 	\\
''		            & ''                 &2023-02-08		& 00010682058	&1883    &S58 	\\
''		            & ''                 &2023-02-11		& 00010682059	&974    &S59	\\
''		            & ''                 &2023-02-13		& 00010682060	&1788    &S60 	\\
''		            & ''                 &2023-02-25		& 00010682061	&842    &S61 	\\
''		            & ''                 &2023-02-27		& 00010682062	&910    &S62 	\\
''		            & ''                 &2023-03-02		& 00010682063	&1780    &S63 	\\
''		            & ''                 &2023-03-04		& 00010682064	&891    &S64 	\\
''		            & ''                 &2023-03-08		& 00010682065	&1800    &S65 	\\
''		            & ''                 &2023-03-11		& 00010682066	&872    &S66 	\\
''		            & ''                 &2023-03-14		& 00010682067	&868    &S67 	\\
''		            & ''                 &2023-03-18		& 00010682068	&935    &S68 	\\
''		            & ''                 &2023-03-20		& 00010682069	&1065    &S69 	\\
''		            & ''                 &2023-03-29		& 00010682071	&809    &S71 	\\
''		            & ''                 &2023-03-30		& 00010682072   &1066    &S72 	\\
''		            & ''                 &2023-04-16		& 00010682073	&888    &S73 	\\
''		            & ''                 &2023-04-22		& 00010682074   &938    &S74 	\\
''		            & ''                 &2023-04-29		& 00010682076	&903    &S76 	\\
''		            & ''                 &2023-05-01		& 00010682077	&1168    &S77 	\\
''		            & ''                 &2023-05-05		& 00010682078	&902    &S78 	\\\hline

\hline

Optical     &{\it HTC}  & 2022-10-31    &  HCT-2022-C3-P10         & 2400\\

\hline


Radio		& {\it VLBA}	&	2022-03-05	& BM527 & 12600 \\
''			& {\it VLBA}	&	2022-08-05	& BY177B &  10080 \\

\hline 
\end{tabular}  

{{\it VLBA}= Very Long Baseline  Array. }
\end{table*}


\section{Observation, data reduction and data analysis}\label{sec:obs}

\subsection{Swift XRT and UVOT}\label{swift_obs}

Observations of 1ES~1927+654 were carried out by {\it The Neil Gehrels Swift Observatory} (\swift{} hereafter) initially at a monthly cadence from January 2022 to November 2022, and then at a weekly and bi-weekly cadence from December 2022 to May 2023 (See Table~\ref{Table:obs}) under a Director's Discretionary Time (DDT) program (PI: S.Laha). The earlier \swift{} XRT~\citep{2005SSRv..120..165B} and UVOT~\citep{2005SSRv..120...95R} observations of this source (prior to 31st Dec 2021) have been reported in our previous work \cite{laha2022}, and we use those results in this work for comparison and to portray a complete picture of the phenomenon. 

We followed the automated XRT analysis approach via the online tools\footnote{https://www.swift.ac.uk/user-objects} \citep{Evans2009_xrt} for the XRT data in all our observations (S33-S78) as recommended for point sources by the \swift{} help desk. Refer to \cite{laha2022} for a full description of UVOT data reprocessing and analysis, which we follow here. The UV flux densities were corrected for Galactic absorption using the correction magnitude of $\rm A_{\lambda}=0.690$ obtained from the NASA Extragalactic Database (NED\footnote{https://ned.ipac.caltech.edu}).

We used a simple absorbed power-law model in XSPEC \citep{1996ASPC..101...17A}  to fit the $0.3-10\kev$ XRT grouped spectra (a minimum of 10 counts per bin) for all the \swift{} observations and added a black-body component to model the soft X-ray excess emission below $2\kev$. We note that adding a {\tt black-body} in the soft X-ray did not improve the fit statistics from observations S33 to S48. However, in the latest observations (S49-S78), where the soft X-ray flux increased notably, a {\tt black-body} component improved the fit-statistics significantly with a typical temperature $kT_{\rm e}\sim 0.2\kev$. Table~\ref{Table:xray_obs} quotes the best-fit parameters and fit statistics.

We measured the UV monochromatic flux using the UVW2 filter for all \swift{} observations and corrected it for Galactic reddening and extinction. We created the UVOT source and background spectral files using the {\it UVOT2PHA} tool and used the response files provided by the \swift{} team.

We estimated the bolometric luminosity ($\lbol$) and hence the accretion rate ($\lambdaedd$) of the source using the relation $\lambdaedd=\lbol/\ledd$, where $\ledd$ is the Eddington rate of the source assuming a black hole of mass $\sim 10^6\msol$ \citep{Ricci2020}. The value is estimated using the galaxy's stellar mass inferred from K-band photometry relation from \cite{Kormendy2013}. Here, we note that the estimates of the black hole mass of 1ES1927+654 varied up to $\sim 10^{7}\msol$ \citep{trak19} where the authors used the virial method of broad Balmer emission lines. However, it is possible that due to the transient nature of the event, the clouds emitting the broad Balmer lines might not have had the time to virialize, as shown by their variable widths \citep{Ricci2020}. To estimate $\lbol$, we have used the simple approach $\lbol=L_{\rm UV}+L_{\rm X-ray}$, where $L_{\rm UV}$ is the integrated UV luminosity in the band $0.001-100\ev$ obtained by fitting the {\it Swift} UVOT photometric data, and $L_{\rm X-ray}$ is the X-ray luminosity in the band $0.3-10\kev$ measured from the best-fit model obtained from the spectral fitting of each observation. The UVOT data (the 3 UV mono-chromatic data points, UVM1, UVW1, and UVW2) for every \swift{} observation (S01-S78) have been fitted with a disk black body (diskbb) in XSPEC. The model in XSPEC reads as $tbabs\times REDDEN\times diskbb$ and the corresponding luminosity $L_{UV}$ is measured using the command $clumin$, over the energy range $0.01\ev$ to $100\ev$.

\begin{table*}
\fontsize{8}{9}
\centering
  \caption{The spectral parameters obtained using {\it Swift} and \xmm{} UV and X-ray observations of 1ES~1927+654. For comparison with the pre-flare values, we keep the \xmm{} observation. \label{Table:xray_obs}} 
  \begin{tabular}{cccccccccccc} \hline\hline 

ID(DD/MM/YY)			&$F_{0.3-2\kev}^{\rm A}$	&$F_{2-10\kev}^{\rm A}$ &$F_{1.5-2.5\kev}^{\rm A}$  & kT (keV) & $\Gamma$ & UV filter& UV flux density$^{\rm B}$ &$\alpha_{\rm OX}$ & $\rm \chi^2/ \chi^2_{\nu}$ \\ \hline  

X1 (20/05/11)	    &   $9.41\pm0.66$   & $3.92\pm0.08$ & $1.64\pm0.02$ &$0.20\pm0.01$               & $2.21_{-0.02}^{+0.02}$ &UVM2	& $1.34\pm0.03$ &1.004 & $185/1.37$  \\ \hline


S33 (22/03/22)	    &   $18.26\pm1.34$   & $4.66\pm0.66$ & $2.49\pm0.22$ & $-$               &$2.67_{-0.09}^{+0.10}$ &UVW2	& $2.00\pm0.09$ &$1.002$ & $119.36/1.03$\\

S34 (23/04/22)	    &   $14.46\pm1.40$  & $4.59\pm0.66$ & $2.25\pm0.20$ & $-$               &$2.55_{-0.14}^{+0.14}$ &UVW2	& $2.04\pm0.09$ &$1.022$ & $75.91/0.90$\\

S35 (20/05/22)	    &   $8.70\pm1.71$   & $2.81\pm1.05$ & $1.37\pm0.30$ & $-$               &$2.54_{-0.24}^{+0.27}$ &UVW2	& $1.83\pm0.08$ &$1.087$ & $27.03/1.13$\\

S36 (26/05/22)	    &  $10.04\pm1.48$   & $5.02\pm1.79$ & $2.12\pm0.36$ & $-$               &$2.26_{-0.22}^{+0.22}$ &UVW2	& $1.91\pm0.12$ &$1.021$ & $34.49/0.84$\\

S37 (20/06/22)	    &   $12.42\pm1.30$   & $4.03\pm0.89$ & $1.96\pm0.26$ & $-$               &$2.54_{-0.14}^{+0.15}$ &UVW2	& $1.91\pm0.09$ &$1.034$ & $47.19/0.73$\\

S38 (20/07/22)	    &   $13.66\pm1.46$   & $5.22\pm1.04$ & $2.37\pm0.29$ & $-$               &$2.45_{-0.13}^{+0.14}$ &UVW2	& $2.04\pm0.09$ &$1.013$ & $68.35/1.04$\\

S39 (20/08/22)	    &   $18.23\pm2.03$   & $6.35\pm1.37$ & $2.99\pm0.39$ & $-$               &$2.50_{-0.14}^{+0.15}$ &UVW2	& $2.28\pm0.13$ &$0.993$ & $35.88/0.61$\\

S40 (24/08/22)	    &   $20.39\pm2.16$   & $5.37\pm1.16$ & $2.83\pm0.37$ & $-$               &$2.66_{-0.14}^{+0.15}$ &UVW2	& $2.02\pm0.09$ &$0.982$ & $59.22/0.90$\\

S41 (26/08/22)	    &   $14.14\pm1.16$   & $4.62\pm0.76$ & $2.24\pm0.22$ & $-$               &$2.54_{-0.11}^{+0.11}$ &UVW2	& $2.24\pm0.11$ &$1.038$ & $99.27/0.97$\\

S42 (26/09/22)	    &   $17.71\pm1.23$   & $4.79\pm0.67$ & $2.50\pm0.21$ & $-$               &$2.64_{-0.09}^{+0.09}$ &UVW2	& $1.98\pm0.09$ &$0.999$ & $115.91/0.92$\\

S43 (28/10/22)	    &   $19.71\pm3.08$   & $6.55\pm2.22$ & $3.15\pm0.62$ & $-$               &$2.52_{-0.22}^{+0.23}$ &UVW2	& $1.96\pm0.10$ &$0.959$ & $45.61/1.26$\\

S44 (20/11/22)	    &   $25.55\pm4.95$   & $6.35\pm2.42$ & $3.43\pm0.79$ & $-$               &$2.69_{-0.24}^{+0.24}$ &UVW2	& $2.02\pm0.12$ &$0.950$ & $33.98/0.85 $\\

S45 (26/11/22)	    &   $26.02\pm2.17$   & $7.18\pm1.15$ & $3.72\pm0.37$ & $-$               &$2.73_{-0.10}^{+0.11}$ &UVW2	& $2.30\pm0.13$ &$0.958$ & $78.79/0.89$\\

S46 (03/12/22)	    &   $42.71\pm5.76$   & $10.52\pm3.24$ & $5.70\pm1.03$ & $-$               &$2.69_{-0.20}^{+0.21}$ &UVW2	& $2.00\pm0.09$ &$0.864$ & $28.10/0.62$\\

S47 (14/12/22)	    &   $26.01\pm1.51$   & $6.14\pm0.71$ & $3.38\pm0.25$ & $-$               &$2.71_{-0.07}^{+0.08}$ &UVW2	& $2.02\pm0.11$ &$0.952$ & $161.52/1.11$\\

S48 (17/12/22)	    &   $25.92\pm1.72$   & $5.41\pm0.70$ & $3.14\pm0.27$ & $-$               &$2.79_{-0.09}^{+0.09}$ &UVW2	& $1.86\pm0.10$ &$0.951$ & $89.07/0.70$\\

S49 (21/12/22)	    &   $34.81\pm2.12$   &$10.12\pm1.84$ & $5.10\pm0.44$ & $0.20\pm 0.05$    &$2.47_{-0.19}^{+0.16}$ &UVW2	& $2.00\pm0.10$ &$0.882$ & $126.62/0.86$\\

S50 (02/01/23)	    &   $27.22\pm3.05$   &$11.54\pm4.85$ & $3.68\pm0.78$ & $0.17\pm 0.04$    &$2.01_{-0.90}^{+0.50}$ &UVW2	& $1.76\pm0.10$ &$0.915$ & $23.36/0.83$\\

S51 (07/01/23)	    &   $39.74\pm3.94$   &$8.01\pm2.49$ & $4.31\pm0.65$ & $0.17\pm 0.04$    &$2.56_{-0.41}^{+0.29}$ &UVW2	& $1.87\pm0.11$ &$0.899$ & $66.94/0.96$\\

S52 (14/01/23)	    &   $32.95\pm1.68$   &$6.42\pm1.06$ & $3.80\pm0.27$ & $0.19\pm 0.03$    &$2.67_{-0.16}^{+0.14}$ &UVW2	& $1.91\pm0.11$ &$0.924$ & $151.81/0.95$\\

S53 (21/01/23)	    &   $30.75\pm2.83$   &$7.41\pm2.47$ & $3.70\pm0.49$ & $0.19\pm 0.03$    &$2.29_{-0.49}^{+0.33}$ &UVW2	& $1.96\pm0.11$ &$0.932$ & $74.32/1.08$\\

S54 (25/01/23)	    &   $37.19\pm2.26$   &$7.89\pm1.51$ & $4.82\pm0.46$ & $0.22\pm 0.05$    &$2.64_{-0.17}^{+0.16}$ &UVW2	& $1.89\pm0.11$ &$0.882$ & $112.22/0.80$\\

S55 (28/01/23)	    &   $42.53\pm2.80$   &$8.30\pm1.79$ & $5.19\pm0.50$ & $0.20\pm 0.02$    &$2.54_{-0.22}^{+0.19}$ &UVW2	& $1.63\pm0.22$ &$0.845$ & $119.84/0.92$\\

S56 (01/02/23)	    &   $34.86\pm2.24$   &$6.12\pm0.99$ & $3.54\pm0.24$ & $0.19\pm 0.03$    &$2.54_{-0.22}^{+0.27}$ &UVW2	& $1.96\pm0.11$ &$0.940$ & $106.9/0.86$\\

S57 (04/02/23)	    &   $35.95\pm3.20$   &$6.22\pm2.19$ & $3.86\pm0.51$ & $0.19\pm 0.05$    &$2.68_{-0.39}^{+0.30}$ &UVW2	& $1.91\pm0.11$ &$0.921$ & $74.79/0.95$\\

S58 (08/02/23)	    &   $35.74\pm1.98$   &$8.82\pm1.39$ & $4.11\pm0.31$ & $0.17\pm 0.02$    &$2.36_{-0.26}^{+0.21}$ &UVW2	& $1.62\pm0.11$ &$0.883$ & $134.25/0.91$\\

S59 (11/02/23)	    &   $38.15\pm3.56$   &$6.54\pm1.82$ & $3.59\pm0.51$ & $0.15\pm 0.03$    &$2.66_{-0.36}^{+0.27}$ &UVW2	& $1.60\pm0.08$ &$0.904$ & $66.18/0.89$\\

S60 (13/02/23)	    &   $31.94\pm2.09$   &$7.71\pm3.20$ &$3.78\pm0.35$ & $0.16\pm 0.03$    &$2.48_{-0.24}^{+0.20}$ &UVW2	& $2.02\pm0.13$ &$0.934$ & $92.94/0.74$\\

S61 (25/02/23)	    &   $37.50\pm4.15$   &$6.83\pm2.44$ &$5.14\pm1.08$ & $0.25\pm 0.06$    &$2.96_{-0.37}^{+0.36}$ &UVW2	& $2.17\pm0.15$ &$0.895$ & $58.62/0.95$\\

S62 (27/02/23)	    &   $42.61\pm3.56$   &$7.18\pm2.31$ &$4.35\pm0.53$ & $0.19\pm 0.03$    &$2.55_{-0.36}^{+0.38}$ &UVW2	& $1.76\pm0.10$ &$0.887$ & $71.96/0.83$\\

S63 (02/03/23)	    &   $38.06\pm2.43$   &$8.31\pm1.85$ &$4.08\pm0.37$ & $0.18\pm 0.02$    &$2.34_{-0.28}^{+0.26}$ &UVW2	& $1.89\pm0.11$ &$0.910$ & $146.87/1.15$\\

S64 (04/03/23)	    &   $44.42\pm3.53$   &$8.40\pm2.15$ &$5.33\pm0.68$ & $0.21\pm 0.05$    &$2.63_{-0.24}^{+0.24}$ &UVW2	& $1.81\pm0.11$ &$0.858$ & $65.41/0.71$\\

S65 (08/03/23)	    &   $33.07\pm2.03$   &$6.60\pm1.36$ &$3.43\pm0.31$ & $0.17\pm 0.02$    &$2.51_{-0.29}^{+0.23}$ &UVW2	& $1.91\pm0.11$ &$0.941$ & $122.24/0.91$\\

S66 (11/03/23)	    &   $42.65\pm4.15$   &$10.90\pm3.15$ &$4.88\pm0.63$ & $0.17\pm 0.03$   &$2.27_{-0.30}^{+0.40}$ &UVW2	& $1.67\pm0.10$ &$0.860$ & $61.13/0.90$\\

S67 (14/03/23)	    &   $39.96\pm3.76$   &$9.35\pm3.11$ &$4.96\pm0.65$ & $0.19\pm 0.05$    &$2.52_{-0.40}^{+0.31}$ &UVW2	& $1.83\pm0.11$ &$0.872$ & $58.54/0.81$\\

S68 (18/03/23)	    &   $41.08\pm3.53$   &$8.52\pm2.64$ &$4.28\pm0.53$ & $0.18\pm 0.02$    &$2.27_{-0.53}^{+0.37}$ &UVW2	& $2.02\pm0.11$ &$0.913$ & $71.32/0.91$\\

S69 (20/03/23)	    &   $43.01\pm5.17$   &$9.53\pm2.97$ &$5.71\pm0.98$ & $0.24\pm 0.06$    &$2.70_{-0.44}^{+0.43}$ &UVW2	& $1.54\pm0.09$ &$0.820$ & $56.39/1.01$\\

S71 (29/03/23)	    &   $36.97\pm3.29$   &$9.87\pm2.95$ &$4.31\pm0.55$ & $0.17\pm 0.02$    &$2.27_{-0.43}^{+0.32}$ &UVW2	& $2.18\pm0.12$ &$0.925$ & $82.06/1.05$\\

S72 (30/03/23)	    &   $39.48\pm3.27$   &$4.49\pm1.67$ &$2.86\pm0.42$ & $0.17\pm 0.02$    &$2.69_{-0.55}^{+0.36}$ &UVW2	& $2.33\pm0.15$ &$1.004$ & $69.94/0.82$\\

S73 (16/04/23)	    &   $36.48\pm2.89$   &$7.00\pm1.93$ &$4.10\pm0.45$ & $0.19\pm 0.05$    &$2.63_{-0.31}^{+0.26}$ &UVW2	& $1.80\pm0.09$ &$0.901$ & $86.08/0.90$\\

S74 (22/04/23)	    &   $42.50\pm3.93$   &$7.95\pm3.00$ &$4.80\pm0.65$ & $0.20\pm 0.04$    &$2.63_{-0.49}^{+0.32}$ &UVW2	& $2.28\pm0.12$ &$0.914$ & $66.04/0.85$\\

S76 (29/04/23)	    &   $46.08\pm6.15$   &$9.00\pm4.87$ &$4.60\pm0.97$ & $0.18\pm 0.04$    &$2.30_{-0.69}^{+0.66}$ &UVW2	& $2.28\pm0.13$ &$0.921$ & $33.75/0.87$\\

S77 (01/05/23)	    &   $40.69\pm3.55$   &$4.32\pm1.44$ &$3.29\pm0.44$ & $0.19\pm 0.03$    &$2.90_{-0.34}^{+0.29}$ &UVW2	& $2.09\pm0.09$ &$0.963$ & $83.03/1.00$\\

S78 (05/05/23)	    &   $41.17\pm3.68$   &$7.07\pm2.41$ &$4.44\pm0.59$ & $0.20\pm 0.04$    &$2.58_{-0.47}^{+0.32}$ &UVW2	& $2.11\pm0.11$ &$0.914$ & $40.89/0.62$\\\hline

\hline 
\end{tabular}  
$^{\rm A}$ Flux in units of $ 10^{-12}\funit{}$ corrected for Galactic absorption.\\
$^{\rm B}$ UV flux density in units of $ 10^{-15}\funita{}$\\
$\alpha_{\rm OX}=-0.385\log(\rm F_{2\kev}/F_{2500\rm \AA})$\\
The UV flux density was corrected for Galactic absorption using the correction magnitude of $\rm A_{\lambda}=0.690$ obtained from NED.

\end{table*}

\subsection{VLBA}

1ES~1927+654 was observed by the Very Long Baseline Array (VLBA) on 05 March 2022 under DDTs proposals BM527 (PI: E. Meyer) and as part of project BY177B (PI: X. Yang) on 05 August 2022. A standard dual-polarization 6 cm frequency setup was used, with central channel frequencies of 4868 MHz, 4900 MHz, 4932 MHz, 4964 MHz, 4996 MHz, 5028 MHz, 5060 MHz, and 5092 MHz, and a total bandwidth of 32 MHz. As the source was expected to be too faint for self-calibration, a relatively fast-switching cadence between the target and a bright calibrator source was used in both projects for phase-referencing \citep{1995ASPC...82..327B}. The observations were $3.5$ hours and $2.8$ hours resulting in acceptable $uv$ coverage for imaging.

The data were checked for radio frequency interference and then calibrated using VLBARUN (Astronomical Image Processing System or AIPS VLBA pipeline; \cite{AIPS}). The imaging and analysis of source structure were done using a custom version of \texttt{DIFMAP} \cite{shep94} \texttt{ngDIFMAP} \citep{royc23}. The best model (as discussed in \citealt{laha2022} for a previous VLBA observation), as determined by the reduced chi-squared, is a point source centered on a uniformly bright disk.  The details of the procedure are as given in \cite{laha2022}, and Table \ref{Table:radio} lists the details of the point source and extended disk that were found to describe the source best. We used Monte Carlo simulations \citep[e.g.,][]{briggs95,chael18,royc23} to verify that the extended emission around the point source is intrinsic to the source and is not an artefact of interferometric errors. In Table \ref{Table:gb}, we list the ratio of the core radio flux density to the $2-10\kev$ flux \citep[also known as the G\"udel Benz relation,][]{guedel93}.

We find that the point source flux density has increased by a factor of $\sim1.8$ in a year from March 2021 to March 2022, followed by a decrease by a similar factor by August 2022. The flux density of the uniform disk has remained mostly unchanged, while its structure has shrunk by $\sim 20\%$ between March 2021 and August 2022.


\begin{table*}
\centering
  \caption{Details of the radio observations, with corresponding flux density measurements.  In cases where only an unresolved core is observed, the total flux density equals the core flux density.  For cases where we detect resolved extended emission, the central point source (PS) flux density is noted alongside the extended flux density and the semi-major and semi-minor axes of a best-fitting uniform disk model. Note that ``flux$"$ in the table headers refers to flux density. The brightness temperature $T_B$ (lower limit) \citep{kov2005tb} has been calculated for the point source at 5 GHz}.\label{Table:radio} 
  \setlength{\tabcolsep}{2.0pt}
  \begin{tabular}{cccccccccc} \hline\hline 

Obs. & Freq. & Date & Total flux & Central PS flux &  Extended flux & Disk Dimensions &  RMS & Resolution & $T_B$ \\
& (GHz) & {\footnotesize (MM/YY)} & {\footnotesize (mJy/beam)} & (mJy) &  (mJy) & (mas$\times$mas) & (Jy/beam) & (mas$\times$mas) & ($\times 10^6$ K) \\
VLBA & 4.98 & 03/22 & 6.6$\pm$0.5 & 2.4$\pm$0.1 &  4.2$\pm$0.4 & 4.4$\pm$0.1$\times$4.2$\pm$0.1 & 1.0$\times10^{-4}$ & 3.42$\times$1.72 & $>9.7$ \\
VLBA & 4.98 & 08/22 & 5.6$\pm$0.5 & 1.5$\pm$0.1 &  4.1$\pm$0.4 & 4.0$\pm$0.1$\times$3.2$\pm$0.1 & 6.8$\times10^{-5}$ & 3.59$\times$1.36 & $>5.7$ \\

\hline 

\end{tabular} 
\end{table*}

 
\begin{table*}
\centering
  \caption{Semi-contemporaneous X-ray and Radio fluxes (either from 1.5 or 5 GHz VLBI) and the Güdel-Benz Relation.}\label{Table:gb}
  \begin{tabular}{cccccccc} \hline\hline 

X-ray Epoch	& Mean 2-10 keV X-ray flux ($F_X$) & VLBI epoch &  VLBI flux ($F_R$) & Ratio of mean fluxes $F_R/F_X$ & \\
(MM/YY) & ($\funit{}$) & (MM/YY) & ($\funit{}$) & &  \\
\hline
 $^*$05/11& $3.70\pm 0.07\times 10^{-12}$ & 08/13 & $5.00\pm0.50\times10^{-16}$ & $1.35\pm0.11\times10^{-4}$ \\
 $^*$05/11& $3.70\pm 0.07\times 10^{-12}$& 03/14 & $1.80\pm0.20\times10^{-16}$ & $4.86\pm0.05\times10^{-5}$ \\
12/18 & $1.70\pm 0.50\times10^{-12}$ & 12/18 & $3.50\pm0.30\times10^{-17}$ & $2.05\pm0.42\times10^{-5}$ \\
03/21 & $4.40\pm 1.04\times10^{-12}$ & 03/21 & $6.40\pm0.50\times10^{-17}$ & $1.45\pm0.23 \times 10^{-5}$ \\\hline
03/22 & $4.66\pm 0.66\times10^{-12}$ & 03/22 & $1.10\pm0.05\times10^{-16}$ & $2.36\pm0.07 \times10^{-5}$ \\
08/22 & $5.80\pm 0.90\times10^{-12}$ & 08/22 & $6.40\pm0.50\times10^{-17}$ & $1.10\pm0.09 \times10^{-5}$ \\
\hline
\end{tabular} 

$^*$ There are no contemporary X-ray observations of this source along with VLBI in 2013 and 2014. Hence we used the $2-10\kev$ flux from the 2011 \xmm{} observation.
\end{table*}

\subsection{Optical spectroscopy} 

\subsubsection{The Himalayan Chandra Telescope observation}
We observed the source on October 31, 2022, using the 2m Himalayan Chandra Telescope (HCT) located at Hanle, India. We used grism 7 covering a wavelength range of 4000\AA\, to 7000\AA\, and 2 arcsec slit for observation. Flux calibration was done using a spectroscopic standard star BD+28 4211. Standard spectroscopic reduction using the Image Reduction and Analysis Facility (IRAF) package is used to extract the spectra, including flat subtraction, wavelength, and flux calibration. 

To estimate the spectral properties of the optical emission line, we first modeled the host galaxy using PPXF code \citep{2017MNRAS.466..798C} and MILES stellar templates \citep{2011A&A...532A..95F}. We subtracted the best-fit stellar model from the spectrum leading to the pure AGN spectrum. Prominent narrow-emission lines such as [O III], N II, [S II] are clearly present in the optical spectrum; however, the H$\beta$ line is only detected in the host-subtracted spectrum. This is similar to our earlier findings in \cite{laha2022}. To estimate the emission line flux and width, we performed a multi-component modeling of the host-subtracted spectrum using Gaussian functions separately for the H$\beta$ and H$\alpha$ regions. We model all the narrow components using a single Gaussian. In the H$\beta$ complex, narrow H$\beta$ and the [O III]4959, 5007 doublets were modeled while keeping the flux ratio fixed at the theoretical values. 
In the H$\alpha$ region, the narrow H$\alpha$, [NII]6549, [NII]6568, [S II]6717, [S II]6732 were modeled where the flux ratio of NII doublets was fixed to their theoretical values, and SII doublets ratio was fixed to unity. During the fit, velocity shift and width were kept as free parameters. The best-fit model was obtained via the chi-square minimization method, and the uncertainty on the model parameter was obtained by the Monte Carlo approach (see \citealt{rakshit2020}).
We noticed excess emission in the residual spectrum around the H$\alpha$ line while modeling with a single Gaussian (narrow component). Therefore, we have added another Gaussian profile with a line full-width-at-half-maximum (FWHM) larger than 900 km/s to represent the broad component for H$\alpha$ regions. We find a statistically improved fit with this addition of the broad component.


\subsubsection{The Lowell Discovery Telescope observation}
We obtained optical spectroscopy on October 30, 2022, using the DeVeny spectrograph mounted on the 4.3m Lowell Discovery Telescope (LDT) in Happy Jack, AZ, for a total exposure of 600 s. DeVeny was configured with the 300 g mm$^{-1}$ grating and a 1.5\arcsec slit width. The spectrum covers wavelengths $3600\, \AA$\,-\,$8000\, \AA$ at a dispersion of $2.2\,\AA$ pix$^{-1}$. The data were processed using methods described in \cite{prochaska2020}, that is, to perform bias subtraction, flat fielding, cosmic ray removal, trace extraction, and telluric corrections, with Ne, Ar, Hg, and Cd arc lamps used for wavelength calibration. The spectrum was flux calibrated using observations of the spectrophotometric standard star G191-B2B obtained on the same night.

 The LDT spectrum does not have a good absolute flux calibration; hence, the spectrum was re-scaled to have the same level as the GTC spectra in 2021, as indicated by the photometry obtained by ZTF (minimal change $<0.03$ magnitude). That is, the r and g magnitudes did not vary by more than $3\%$. First, the spectrum was corrected of galactic interstellar extinction. The stellar population contribution was subtracted using pPXF. The line measurements from the LDT observation are reported in Table \ref{Table:optical_emlines_Lowell} and are consistent with the results of HCT.


\begin{table*}
\centering
  \caption{Optical emission line properties without (the values with broad components are given in the parenthesis) the broad component. }\label{Table:optical_spectral_prop}
  \begin{tabular}{ccccc} \hline\hline 
Lines               &    Center wavelength               &  FWHM                &  Flux \\
                    &     (\AA)                          &  (km s$^{-1}$)       & (erg s$^{-1}$ cm$^{-2}$)\\
                    &                                    &                      &    $\times 10^{-15}$      \\\hline
H$\beta$(NC)        &    4857.5$\pm$3.9                  &  899 $\pm$ 100       & 3.67 $\pm$ 0.30  \\
$\rm [OIII]4959$    & 4953.6$\pm$3.9                     & 706$\pm$71           & 3.78$\pm$0.10 \\
$\rm [OIII]5007$    & 5001.4$\pm$3.9                     & 721$\pm$23           & 11.42$\pm$0.30 \\
$\rm [NII]6549 $    & 6548.5$\pm$4.1 (6551.17$\pm$4.14)  & 421$\pm$63 (673$\pm$170)   & 0.72$\pm$0.09 (0.50$\pm$0.18)\\
H$\alpha$(BC)       &     (6563.5$\pm$5.5)               &  (1172$\pm$327)      & (5.06$\pm$1.30)  \\ 
H$\alpha$(NC)       & 6564.0$\pm$4.1 (6564.8$\pm$4.1)    & 673$\pm$ 63 (389$\pm$66)   & 5.94$\pm$0.42 (1.98$\pm$0.91)\\
$\rm [NII]6585$     & 6585.5$\pm$4.1 (6587$\pm$4.1)      & 640$\pm$ 106 (491$\pm$149) & 2.23$\pm$0.30 (1.48$\pm$0.53) \\
$\rm [SII]6718 $    & 6718.1$\pm$4.1                     & 478$\pm$ 113         & 0.68$\pm$0.08 \\
$\rm [SII]6732 $    & 6727.8$\pm$4.1                     & 520$\pm$ 133         & 0.69$\pm$0.09 \\

\hline
\end{tabular}   
\end{table*}

\begin{table*}
\centering
  \caption{ Optical emission line properties measured in the 
  the Lowell (LDT) spectrum.}\label{Table:optical_emlines_Lowell}


\begin{tabular}{lccccl} \hline\hline
Line ID & Center Wave. & Flux$_{obs}$ & Flux$_{dered}$ & FWHM &  $\rm Comments^{a}$ \\
            &     (\AA)      & (erg s$^{-1}$ cm$^{-2}$) & (erg s$^{-1}$ cm$^{-2}$) &  (km s$^{-1}$) \\
            &          & $\times 10^{-15}$ & $\times 10^{-15}$ &    \\ \hline
$\rm [OII]3727$ & 3727.49$\pm$0.26 & 2.52$\pm$0.28 & 3.48$\pm$0.39 & 710.6$\pm$60.4 & higher (48\%) \\
HeII4686 & 4687.99$\pm$0.85 & 0.74$\pm$0.19 & 0.96$\pm$0.25 & 676.1$\pm$135.2   & same \\
Hb & 4862.10$\pm$0.72 & 1.25$\pm$0.29 & 1.61$\pm$0.37 & 821.6$\pm$153.0  & same  \\
$\rm [OIII]4959$ & 4960.92$\pm$0.06 & 2.33$\pm$0.05 & 2.97$\pm$0.07 & 556.5$\pm$9.4  & higher (52\%) \\
$\rm [OIII]5007$ & 5008.87$\pm$0.06 & 7.07$\pm$0.16 & 9.00$\pm$0.20 & 556.5$\pm$9.4  & higher (52\%) \\
$\rm [NII]6548$ & 6550.10$\pm$0.26 & 0.57$\pm$0.07 & 0.68$\pm$0.08 & 390.5$\pm$41.4  & same \\
Ha & 6565.23$\pm$0.15 & 5.13$\pm$0.21 & 6.03$\pm$0.25 & 537.3$\pm$17.2  & same \\
$\rm [NII]6584$ & 6585.51$\pm$0.26 & 1.73$\pm$0.22 & 2.04$\pm$0.26 & 390.5$\pm$41.4  & same \\
$\rm [SII]6716$ & 6718.03$\pm$1.81 & 0.30$\pm$0.11 & 0.35$\pm$0.13 & 374.3$\pm$75.0  &  same\\
$\rm [SII]6731$ & 6735.27$\pm$0.77 & 0.56$\pm$0.15 & 0.65$\pm$0.18 & 374.3$\pm$75.0  &  same \\
\hline
\end{tabular}\\
$^{\rm a}$ compared to 2021 Gran Telescopio CANARIAS (GTC) observation \citep{laha2022}. \\
\end{table*}


\begin{figure*}
    \centering
    \includegraphics[width=17cm,height=21cm]{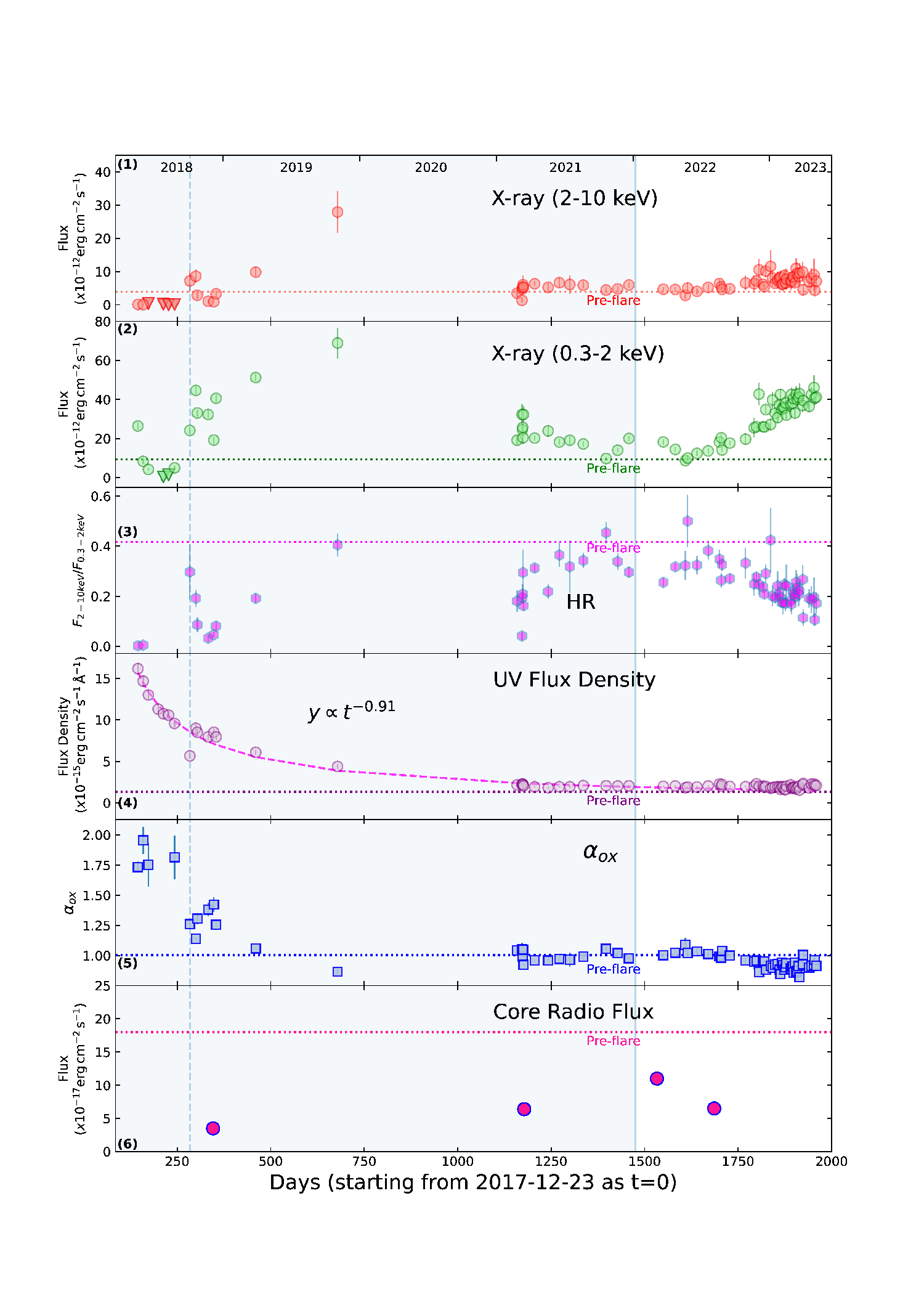}
    \caption{ The light curves of the X-ray, UV, and Radio parameters of the central engine of the AGN 1ES1927+654, as observed by {\it Swift} and VLBA (see Tables \ref{Table:xray_obs} and Table \ref{Table:radio} for details). The shaded region corresponds to the observations discussed in our earlier work \cite{laha2022}, but we have included them here for the sake of completeness and to track the long-term nature of this source. The rest of the figure refers to the new observations used in this work. The start date of the light curve is 2017-12-23, corresponding to the burst date reported by \cite{trak19}. The X-axis is in the units of days elapsed from the start date. The dotted horizontal lines in every panel refer to their pre-flare values (in 2011). The inverted triangles are the upper limits. The X-ray flux in units of $ 10^{-12}\funit{}$ corrected for Galactic absorption. From top to bottom are panels: (1) The $2-10\kev$ X-ray flux, (2) The $0.3-2\kev$ X-ray flux, (3) The hardness ratio: $F_{2-10\kev}/F_{0.3-2\kev}$, (4) The UV (UVW2) flux density (in units of $10^{-15}\funita$), (5) the $\alpha_{\rm OX}$, and (6) the Core radio flux ($<1\pc$ spatial resolution) NOTE: The vertical line corresponds to the observation S8, where the X-ray corona jumps back (created) after being destroyed, and there is a dip in the UV flux by a factor of 2, and also the X-ray spectra become harder (panel 3).} 
    \label{fig:xray_uv_alpha_ox}
\end{figure*}


\begin{figure*}
    \centering
    \includegraphics[width=17cm,height=21cm]{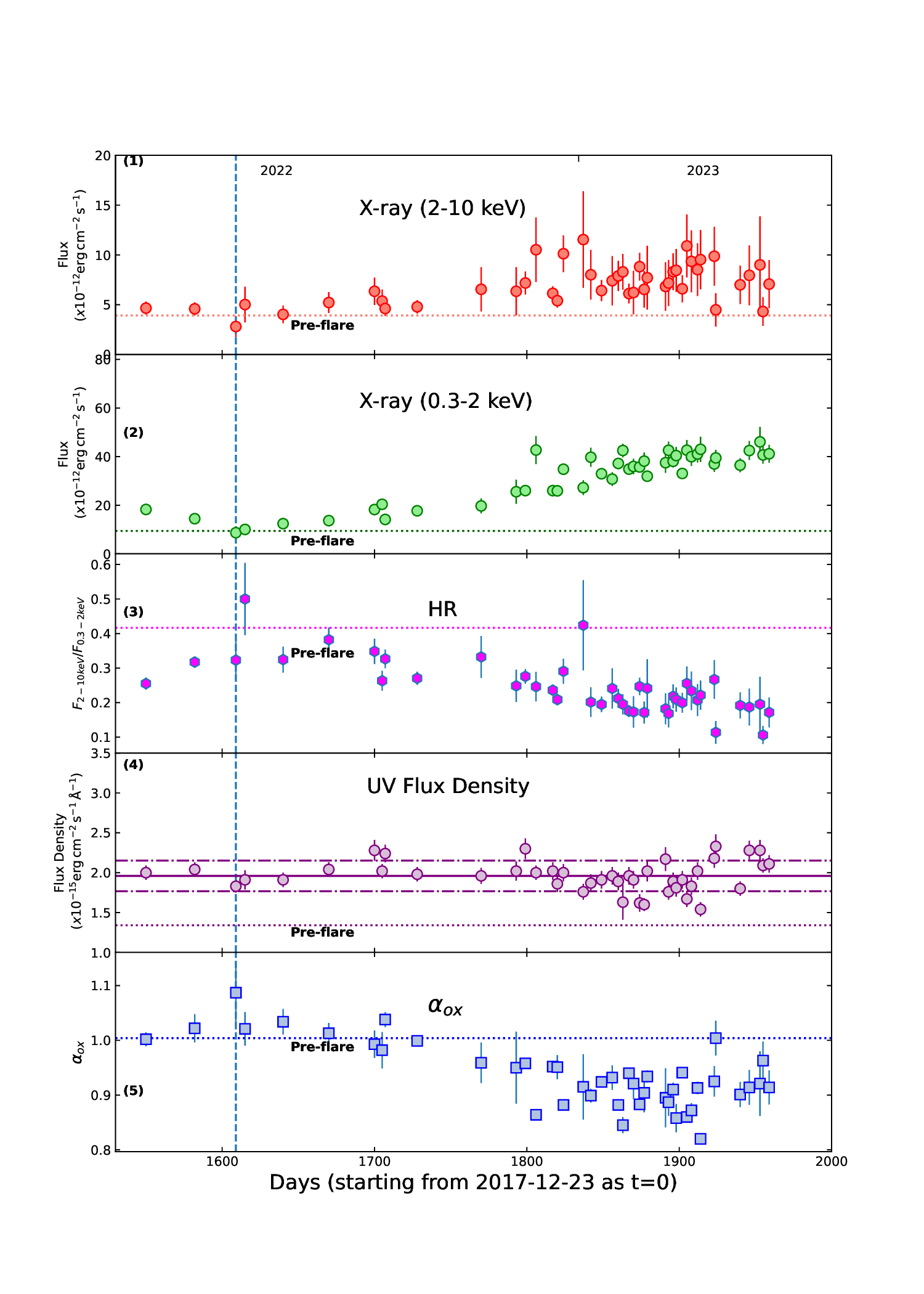}
    \caption{ The zoomed-in version of the X-ray and UV light curves plotted in Fig \ref{fig:xray_uv_alpha_ox} encompassing the observations S33-S78 to highlight the rise of the bright-soft-state of the source 1ES1927+654. The vertical dashed blue line corresponds to observation S35 when both the soft and the hard X-ray fluxes exactly matched their pre-flare values in 2011. See Table \ref{Table:xray_obs} for details. The start date (day zero) of the light curve is 2017-12-23, corresponding to the burst date reported by \cite{trak19}, and the X-axis is in the units of days elapsed from the start date. The dotted horizontal lines in every panel refer to their pre-flare values (in 2011). From top to bottom are panels: (1) The X-ray $2-10\kev$ flux (in units of $10^{-12}\funit$), (2) The X-ray $0.3-2\kev$ flux (in units of $10^{-12}\funit$), (3) The hardness ratio: $F_{2-10\kev}/F_{0.3-2\kev}$, (4)The evolution of the UV (UVW2) flux density along with the mean (1.96) denoted by a solid line and standard deviations (0.19) denoted by dashed-dotted lines (in units of $10^{-15}\funita$), (5) the $\alpha_{\rm OX}$ values during observations S33-S78 are plotted.} 
    \label{fig:xray_uv_alpha_ox_zoomed}
\end{figure*}


\section{Results}\label{sec:results}

In our earlier work \cite{laha2022}, we reported the `initial flare' (May 2018 to December 2021) of the source 1ES~1927+654. Here we report the results from our continuing multi-wavelength campaign from January 2022 to March 2023. We detect a significant rise in the soft X-ray flux (by a factor of 5 in a matter of $\sim 11$ months), which we call the bright-soft-state. Figure \ref{fig:xray_uv_alpha_ox} shows the time evolution of the fluxes in $0.3-2\kev$, $2-10\kev$, UVW2 flux density and the core radio flux. In Figure \ref{fig:xray_uv_alpha_ox_zoomed}, we have zoomed into the time period when the soft X-ray flux has shown a continuous rise. Below we discuss the results obtained from our continuing multi-wavelength observational campaign. 



\subsection{Evolution of $0.3-2\kev$, $2-10\kev$ X-ray fluxes, the UV flux density, and the photon index }

 The soft X-ray flux showed considerable variability in months to week timescale in its post-flare state (after Oct 2021, when it reached the pre-flare state for the first time). It showed a rise and then a drop after that. On 20th May 2022, it reached its pre-flare value of $\sim 9\times 10^{-12}\funit$ again. After this date, the flux showed a steady rise (See Fig \ref{fig:xray_uv_alpha_ox_zoomed} second panel) for eleven straight months. Nevertheless, the biweekly \swift{} observations have captured significant variability at days time-scale. As of May 5th, 2023, the soft X-ray flux level is $\sim 5$ times that of its pre-flare value.

 The hard X-ray flux had reached its pre-flare value much earlier than the soft flux and had shown some variability throughout our campaign. But we do not detect any significant continuous rise of the flux above the pre-flare value like the soft. From Fig \ref{fig:xray_uv_alpha_ox_zoomed} top panel we note that the $2-10\kev$ flux showed a rise (by a factor of $\sim 2$) in January 2023. 
 We note that the flux variation and the overall increase in the soft band over the last year, is more significant compared to the hard band (See Fig~\ref{fig:xray_uv_alpha_ox})

The UVW2 flux density of 1ES~1927+654 was monitored using the \swift{} UVOT, and we quote the UVW2 band for consistency with the literature (although we use all the wavelength bands when we estimate $\lbol$). The UVW2 flux density shows minimal variations in the post-flare state, in particular during the soft X-ray rise from 20th May 2022. The UV flux density values show a $\le 30\%$ variability over a year-long timescale with a mean value of $1.96\times 10^{-15}$ and a standard deviation of $0.19\times 10^{-15}$ (in units of $\funita$) respectively and are shown in the fourth panel (from top) of Figure~\ref{fig:xray_uv_alpha_ox_zoomed}. Here, we also note that the UVW2 flux level has not yet reached the pre-flare value of $1.34\pm 0.03 \times 10^{-15} \funita$. See Table~\ref{Table:xray_obs} for details. As discussed earlier, we estimated the UV band luminosity in the range $(0.01-100)\ev$ using the multi-band photometric data from \swift{}-UVOT. 


After the coronal reappearance, when the source reached its pre-flare state, the power-law photon index $\Gamma$ has been more or less stable around the pre-flare value ($\Gamma \sim 2.2-2.6$) and remained consistent within $3\sigma$ errors, even during the current bright-soft-state.


\subsection{Evolution of $\lambdaedd$}


We find that the Eddington ratio ($\lambdaedd=\lbol/\ledd$) for this source was as high as $\lambdaedd\sim 0.53$ (five times that of the pre-flare state) when the X-ray flux was the highest during the `initial flare' in Nov 2019 (See Table \ref{Table:lambdaedd}), and then it dropped to the pre-flare value ($\lambdaedd\sim 0.09$) and stayed the same until the soft X-ray flux started to rise in May 2022. Over the period of months, we see a gradual rise in the Eddington ratio with the current value of $\lambdaedd\sim 0.25$, half of the bright flux state in the `initial flare'. This increase in the Eddington ratio is solely due to the rise in the soft X-ray flux and not due to the contribution from UV.

\begin{figure}
    \includegraphics[width=9cm]{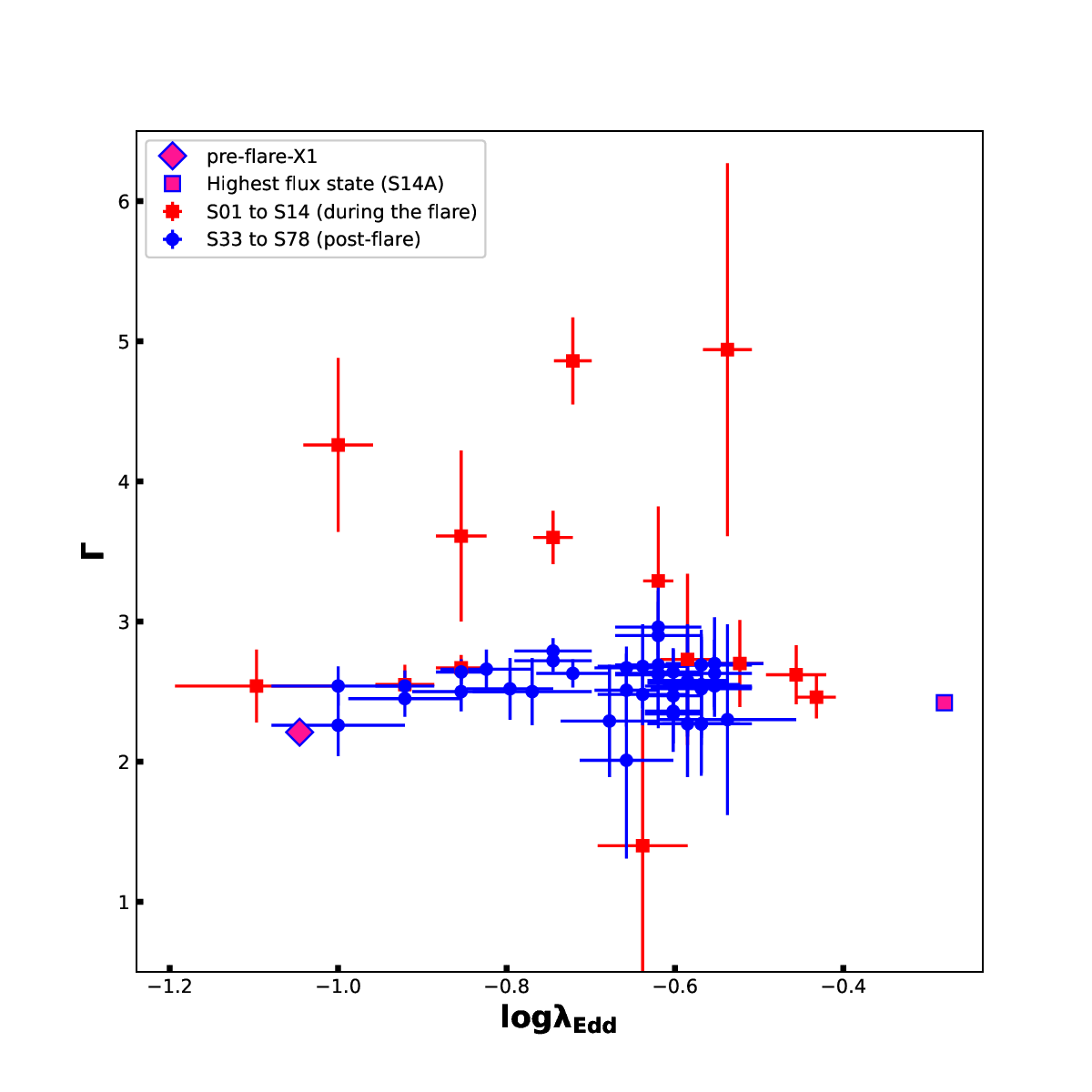}
    \caption{ The evolution of $\log(\lambdaedd)$ and the power-law slope $\Gamma$ of the changing-look AGN 1ES~1927+654 during the initial flare (red squares) and the post-flare state (blue circles). See Table~\ref{Table:lambdaedd} for details). We do not see any correlation between these parameters either during the initial flare or the post-flare states.} 

    \label{fig:gamma_vs_eddington}
\end{figure}

Figure~\ref{fig:gamma_vs_eddington} shows the relation between $\lambdaedd$ and the $\Gamma$. We do not see any strong correlation between these two parameters if we consider the observations during the post-flare bright-soft-states (S33-S78). We find that the $\Gamma$ varies within a narrow range of values with a considerable spread in $\lambdaedd$.



\small   
\setlength{\tabcolsep}{1.4pt}
\begin{longtable}{|c|c|c|c|c|c|}

\caption{The evolution of the disk temperature and $\lambdaedd$ estimated from the {\it Swift} and \xmm{} UV and X-ray observations of 1ES~1927+654. The Eddington rate for an SMBH with a mass of $10^{6}\msol$ is $1.3\times10^{44}\lunit$. 
The X-ray luminosity is calculated from the $0.3-10\kev$ energy band by adding the soft and hard X-ray flux. During the lowest luminosity phase (S03-S07), we assumed an upper limit of $10^{41}\lunit$ for the X-ray flux.}\label{Table:lambdaedd} 


    \\\hline \multicolumn{1}{|c|}{Obs} & \multicolumn{1}{c|}{Diskbb} & \multicolumn{1}{c|}{$L_{UV}$} & \multicolumn{1}{c|}{$L_{X}$} & \multicolumn{1}{c|}{$L_{Bol}$} & \multicolumn{1}{c|}{$\lambdaedd$}\\\hline
    \multicolumn{1}{|c|}{Id} & \multicolumn{1}{c|}{in $\ev$} & \multicolumn{1}{c|}{$10^{43}\lunit$} & \multicolumn{1}{c|}{$10^{43}\lunit$} & \multicolumn{1}{c|}{$L_{UV}+L_{X}$} & \multicolumn{1}{c|}{$L_{Bol}/L_{Edd}$}\\\hline
    
    \endfirsthead
    \multicolumn{6}{c}%
    {{\bfseries \tablename \thetable{} -- continued from previous page}}\\
    \hline \multicolumn{1}{|c|}{Obs} & \multicolumn{1}{c|}{Diskbb} & \multicolumn{1}{c|}{$L_{UV}$} & \multicolumn{1}{c|}{$L_{X}$} & \multicolumn{1}{c|}{$L_{Bol}$} & \multicolumn{1}{c|}{$\lambdaedd$}\\\hline
    \multicolumn{1}{|c|}{Id} & \multicolumn{1}{c|}{in $\ev$} & \multicolumn{1}{c|}{$10^{43}\lunit$} & \multicolumn{1}{c|}{$10^{43}\lunit$} & \multicolumn{1}{c|}{$L_{UV}+L_{X}$} & \multicolumn{1}{c|}{$L_{Bol}/L_{Edd}$}\\\hline

    \endhead

    \hline \multicolumn{6}{|r|}{Continued to the next page} \\ \hline 

    \endfoot

    \hline
    \endlastfoot

X1     &  $03\pm01$   &  $0.28_{-0.05}^{+0.05}$  & $0.86_{-0.01}^{+0.01}$  & $1.14_{-0.06}^{+0.06}$ &  $0.09\pm0.01$ \\\hline
S01    &  $08\pm01$   &  $2.10_{-0.07}^{+0.07}$  & $1.70_{-0.14}^{+0.14}$  & $3.80_{-0.21}^{+0.21}$ &  $0.29\pm0.02$ \\
S02    &  $08\pm02$   &  $1.92_{-0.07}^{+0.08}$  & $0.54_{-0.09}^{+0.09}$  & $2.46_{-0.16}^{+0.17}$ &  $0.19\pm0.01$  \\
S03    &  $06\pm02$   &  $1.81_{-0.06}^{+0.05}$  & $0.01_{-0.01}^{+0.01}$  & $1.82_{-0.07}^{+0.06}$ &  $0.14\pm0.01$ \\
S04    &  $08\pm01$   &  $1.98_{-0.04}^{+0.05}$  & $0.01_{-0.01}^{+0.01}$  & $1.99_{-0.05}^{+0.06}$ &  $0.15\pm0.01$ \\
S05    &  $04\pm02$   &  $1.56_{-0.05}^{+0.05}$  & $0.01_{-0.01}^{+0.01}$  & $1.57_{-0.06}^{+0.06}$ &  $0.12\pm0.01$  \\
S06    &  $08\pm03$   &  $1.38_{-0.05}^{+0.06}$  & $0.01_{-0.01}^{+0.01}$  & $1.39_{-0.06}^{+0.07}$ &  $0.11\pm0.01$   \\
S07    &  $07\pm04$   &  $1.28_{-0.06}^{+0.05}$  & $0.01_{-0.01}^{+0.01}$  & $1.29_{-0.07}^{+0.06}$ &  $0.10\pm0.01$   \\
S08    &  $03\pm02$   &  $1.00_{-0.07}^{+0.09}$  & $2.01_{-0.30}^{+0.30}$  & $3.01_{-0.37}^{+0.39}$ &  $0.23\pm0.03$   \\
S09    &  $10\pm04$   &  $1.11_{-0.04}^{+0.05}$  & $3.42_{-0.28}^{+0.28}$  & $4.53_{-0.32}^{+0.33}$ &  $0.35\pm0.03$   \\
S10    &  $13\pm07$   &  $1.08_{-0.04}^{+0.04}$  & $2.31_{-0.18}^{+0.18}$  & $3.39_{-0.22}^{+0.22}$ &  $0.26\pm0.02$   \\
S11    &  $12\pm06$   &  $0.99_{-0.02}^{+0.03}$  & $2.15_{-0.16}^{+0.16}$  & $3.16_{-0.18}^{+0.19}$ &  $0.24\pm0.01$   \\
S12    &  $12\pm06$   &  $1.06_{-0.03}^{+0.03}$  & $1.29_{-0.12}^{+0.12}$  & $2.35_{-0.15}^{+0.15}$ &  $0.18\pm0.01$   \\
S13    &  $08\pm03$   &  $1.03_{-0.03}^{+0.03}$  & $2.82_{-0.18}^{+0.18}$  & $3.85_{-0.21}^{+0.21}$ &  $0.30\pm0.02$   \\
S14    &  $07\pm04$   &  $0.82_{-0.04}^{+0.04}$  & $3.92_{-0.20}^{+0.18}$  & $4.74_{-0.24}^{+0.22}$ &  $0.37\pm0.02$   \\
S14A   &  $07\pm04$   &  $0.60_{-0.05}^{+0.05}$  & $6.23_{-0.92}^{+0.88}$  & $6.83_{-0.97}^{+0.93}$ &  $0.53\pm0.07$   \\\hline
S33    &  $05\pm02$   &  $0.28_{-0.02}^{+0.03}$  & $1.47_{-0.13}^{+0.13}$  & $1.75_{-0.16}^{+0.15}$ &  $0.14\pm0.01$  \\
S34    &  $04\pm03$   &  $0.30_{-0.02}^{+0.03}$  & $1.23_{-0.13}^{+0.13}$  & $1.53_{-0.16}^{+0.15}$ &  $0.12\pm0.01$  \\
S35    &  $05\pm03$   &  $0.27_{-0.04}^{+0.05}$  & $0.74_{-0.18}^{+0.18}$  & $1.01_{-0.23}^{+0.22}$ &  $0.08\pm0.02$  \\
S36    &  $06\pm02$   &  $0.28_{-0.03}^{+0.03}$  & $0.97_{-0.20}^{+0.20}$  & $1.25_{-0.23}^{+0.22}$ &  $0.10\pm0.02$   \\
S37    &  $04\pm03$   &  $0.29_{-0.03}^{+0.04}$  & $1.06_{-0.14}^{+0.14}$  & $1.35_{-0.17}^{+0.18}$ &  $0.10\pm0.02$ \\
S38    &  $05\pm03$   &  $0.31_{-0.03}^{+0.03}$  & $1.21_{-0.16}^{+0.16}$  & $1.52_{-0.18}^{+0.19}$ &  $0.12\pm0.02$ \\
S39    &  $08\pm05$   &  $0.29_{-0.03}^{+0.03}$  & $1.58_{-0.22}^{+0.22}$  & $1.87_{-0.24}^{+0.25}$ &  $0.14\pm0.02$ \\
S40    &  $05\pm02$   &  $0.28_{-0.05}^{+0.05}$  & $1.66_{-0.21}^{+0.21}$  & $1.94_{-0.25}^{+0.26}$ &  $0.15\pm0.02$ \\
S41    &  $08\pm04$   &  $0.28_{-0.03}^{+0.03}$  & $1.21_{-0.12}^{+0.12}$  & $1.49_{-0.14}^{+0.15}$ &  $0.12\pm0.01$ \\
S42    &  $03\pm01$   &  $0.32_{-0.03}^{+0.02}$  & $1.45_{-0.12}^{+0.12}$  & $1.77_{-0.15}^{+0.14}$ &  $0.14\pm0.01$ \\
S43    &  $03\pm01$   &  $0.33_{-0.03}^{+0.03}$  & $1.69_{-0.29}^{+0.30}$  & $2.02_{-0.32}^{+0.33}$ &  $0.16\pm0.02$ \\
S44    &  $04\pm02$   &  $0.29_{-0.04}^{+0.04}$  & $1.92_{-0.35}^{+0.37}$  & $2.21_{-0.39}^{+0.41}$ &  $0.17\pm0.03$ \\
S45    &  $03\pm02$   &  $0.33_{-0.05}^{+0.05}$  & $2.14_{-0.21}^{+0.20}$  & $2.47_{-0.26}^{+0.25}$ &  $0.19\pm0.02$ \\
S46    &  $05\pm03$   &  $0.30_{-0.03}^{+0.04}$  & $3.22_{-0.55}^{+0.54}$  & $3.52_{-0.58}^{+0.58}$ &  $0.27\pm0.04$ \\
S47    &  $03\pm01$   &  $0.30_{-0.04}^{+0.04}$  & $2.07_{-0.13}^{+0.14}$  & $2.37_{-0.17}^{+0.18}$ &  $0.18\pm0.02$ \\
S48    &  $03\pm01$   &  $0.30_{-0.04}^{+0.05}$  & $2.01_{-0.16}^{+0.16}$  & $2.31_{-0.20}^{+0.21}$ &  $0.18\pm0.02$   \\
S49    &  $03\pm02$   &  $0.37_{-0.05}^{+0.05}$  & $2.89_{-0.24}^{+0.23}$  & $3.26_{-0.29}^{+0.28}$ &  $0.25\pm0.02$ \\
S50    &  $04\pm02$   &  $0.30_{-0.05}^{+0.07}$  & $2.49_{-0.40}^{+0.40}$  & $2.79_{-0.45}^{+0.47}$ &  $0.22\pm0.03$ \\
S51    &  $05\pm03$   &  $0.27_{-0.04}^{+0.06}$  & $3.07_{-0.41}^{+0.40}$  & $3.34_{-0.45}^{+0.45}$ &  $0.26\pm0.03$ \\
S52    &  $08\pm04$   &  $0.27_{-0.05}^{+0.05}$  & $2.53_{-0.17}^{+0.16}$  & $2.80_{-0.22}^{+0.21}$ &  $0.22\pm0.02$ \\
S53    &  $06\pm03$   &  $0.27_{-0.03}^{+0.05}$  & $2.45_{-0.34}^{+0.34}$  & $2.72_{-0.37}^{+0.39}$ &  $0.21\pm0.03$ \\
S54    &  $03\pm02$   &  $0.30_{-0.05}^{+0.05}$  & $2.90_{-0.24}^{+0.24}$  & $3.20_{-0.29}^{+0.29}$ &  $0.25\pm0.02$ \\
S55    &  $04\pm02$   &  $0.30_{-0.05}^{+0.05}$  & $3.39_{-0.31}^{+0.29}$  & $3.69_{-0.36}^{+0.34}$ &  $0.28\pm0.03$ \\
S56    &  $03\pm01$   &  $0.28_{-0.03}^{+0.03}$  & $2.37_{-0.17}^{+0.15}$  & $2.65_{-0.20}^{+0.18}$ &  $0.20\pm0.02$ \\
S57    &  $03\pm01$   &  $0.33_{-0.05}^{+0.05}$  & $2.71_{-0.35}^{+0.36}$  & $3.04_{-0.40}^{+0.41}$ &  $0.23\pm0.03$ \\
S58    &  $03\pm01$   &  $0.32_{-0.06}^{+0.08}$  & $2.87_{-0.24}^{+0.22}$  & $3.19_{-0.30}^{+0.30}$ &  $0.25\pm0.02$ \\
S59    &  $02\pm01$   &  $0.30_{-0.05}^{+0.06}$  & $2.87_{-0.34}^{+0.36}$  & $3.17_{-0.39}^{+0.42}$ &  $0.24\pm0.03$ \\
S60    &  $04\pm02$   &  $0.44_{-0.04}^{+0.04}$  & $2.55_{-0.32}^{+0.34}$  & $2.99_{-0.36}^{+0.38}$ &  $0.23\pm0.03$ \\
S61    &  $06\pm03$   &  $0.30_{-0.03}^{+0.04}$  & $2.85_{-0.42}^{+0.40}$  & $3.15_{-0.45}^{+0.44}$ &  $0.24\pm0.03$ \\
S62    &  $03\pm01$   &  $0.32_{-0.06}^{+0.06}$  & $3.20_{-0.36}^{+0.38}$  & $3.52_{-0.42}^{+0.44}$ &  $0.27\pm0.03$ \\
S63    &  $04\pm02$   &  $0.30_{-0.03}^{+0.04}$  & $2.98_{-0.27}^{+0.29}$  & $3.28_{-0.30}^{+0.33}$ &  $0.25\pm0.02$ \\
S64    &  $03\pm02$   &  $0.28_{-0.04}^{+0.04}$  & $3.40_{-0.38}^{+0.36}$  & $3.68_{-0.42}^{+0.40}$ &  $0.28\pm0.03$ \\
S65    &  $04\pm01$   &  $0.31_{-0.04}^{+0.03}$  & $2.55_{-0.22}^{+0.22}$  & $2.86_{-0.25}^{+0.26}$ &  $0.22\pm0.02$ \\
S66    &  $02\pm01$   &  $0.32_{-0.05}^{+0.06}$  & $3.24_{-0.37}^{+0.37}$  & $3.56_{-0.43}^{+0.42}$ &  $0.27\pm0.03$ \\
S67    &  $05\pm03$   &  $0.28_{-0.04}^{+0.05}$  & $3.17_{-0.44}^{+0.44}$  & $3.55_{-0.48}^{+0.49}$ &  $0.27\pm0.04$ \\
S68    &  $04\pm02$   &  $0.30_{-0.05}^{+0.05}$  & $3.19_{-0.40}^{+0.40}$  & $3.49_{-0.45}^{+0.45}$ &  $0.27\pm0.04$ \\
S69    &  $02\pm02$   &  $0.29_{-0.04}^{+0.05}$  & $3.38_{-0.52}^{+0.50}$  & $3.67_{-0.56}^{+0.55}$ &  $0.28\pm0.04$ \\
S71    &  $05\pm03$   &  $0.33_{-0.04}^{+0.05}$  & $3.01_{-0.39}^{+0.41}$  & $3.31_{-0.43}^{+0.46}$ &  $0.26\pm0.03$ \\
S72    &  $03\pm01$   &  $0.34_{-0.04}^{+0.05}$  & $2.83_{-0.30}^{+0.32}$  & $3.15_{-0.34}^{+0.37}$ &  $0.24\pm0.03$ \\
S73    &  $03\pm01$   &  $0.29_{-0.04}^{+0.05}$  & $2.80_{-0.32}^{+0.30}$  & $3.07_{-0.36}^{+0.35}$ &  $0.24\pm0.03$ \\
S74    &  $08\pm04$   &  $0.28_{-0.01}^{+0.02}$  & $3.24_{-0.44}^{+0.46}$  & $3.50_{-0.45}^{+0.48}$ &  $0.27\pm0.04$ \\
S76    &  $05\pm02$   &  $0.30_{-0.03}^{+0.04}$  & $3.54_{-0.70}^{+0.72}$  & $3.81_{-0.73}^{+0.76}$ &  $0.29\pm0.06$ \\
S77    &  $04\pm02$   &  $0.30_{-0.02}^{+0.02}$  & $2.89_{-0.32}^{+0.30}$  & $3.17_{-0.34}^{+0.32}$ &  $0.24\pm0.03$ \\
S78    &  $04\pm02$   &  $0.33_{-0.04}^{+0.04}$  & $3.10_{-0.40}^{+0.38}$  & $3.40_{-0.44}^{+0.42}$ &  $0.26\pm0.03$ \\
\hline

\end{longtable}
\normalsize


\begin{figure*}
    \includegraphics[width=8.5cm]{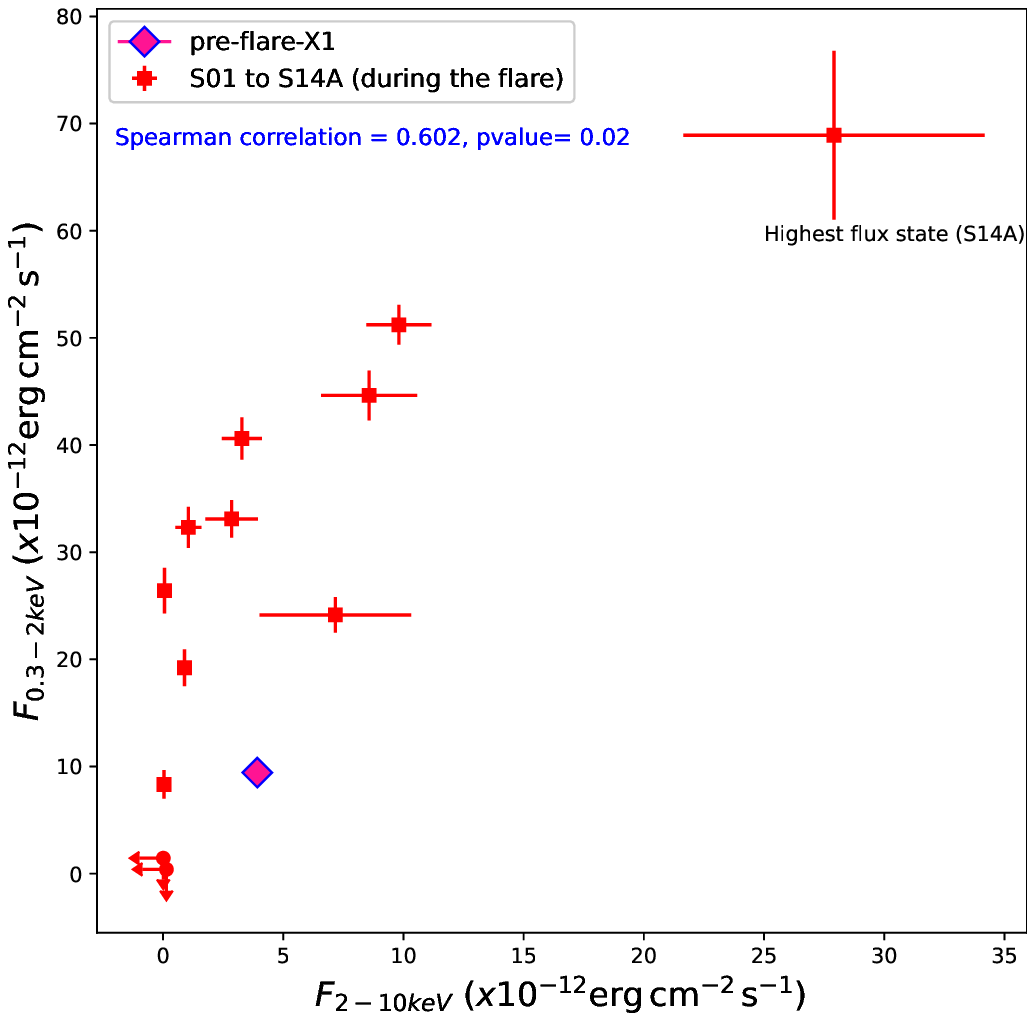}
     \includegraphics[width=8.5cm]{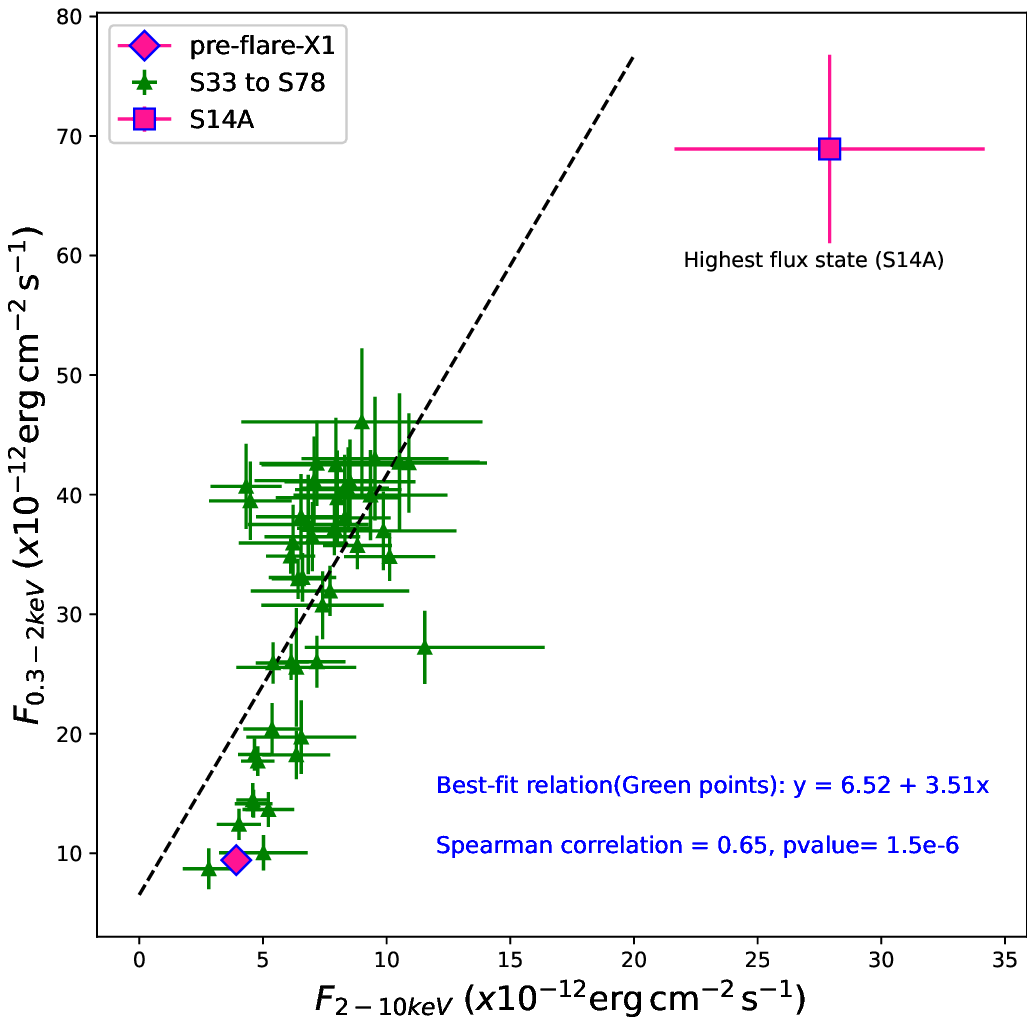}
    
    \caption{ {\it Left:} The relation between the soft X-ray ($0.3-2\kev$) and hard X-ray ($2-10\kev$) fluxes during the initial flare. The red circles denote the X-ray flaring period covering 2018-2019 (observations S01-S14A). The upper limits denote the lowest state in X-rays in both axes in the lower left corner of the figure. The soft and the hard X-rays do not show any significant correlation. {Right:} Same as left, but for the new observations (S33-S78) encompassing the bright-soft-state shows a significant correlation between the soft and hard X-ray solely driven by the power-law flux alone due to the soft nature of the source. The best fit linear regression fit gives us: $y=6.47+3.52x$, with a Spearman rank correlation coefficient=$0.65$, and a statistical significance $> 99.99\%$ 
    .}
           \label{fig:soft_hard_correlation}
\end{figure*}

\begin{figure*}
    \centering
     \includegraphics[width=5.9cm]{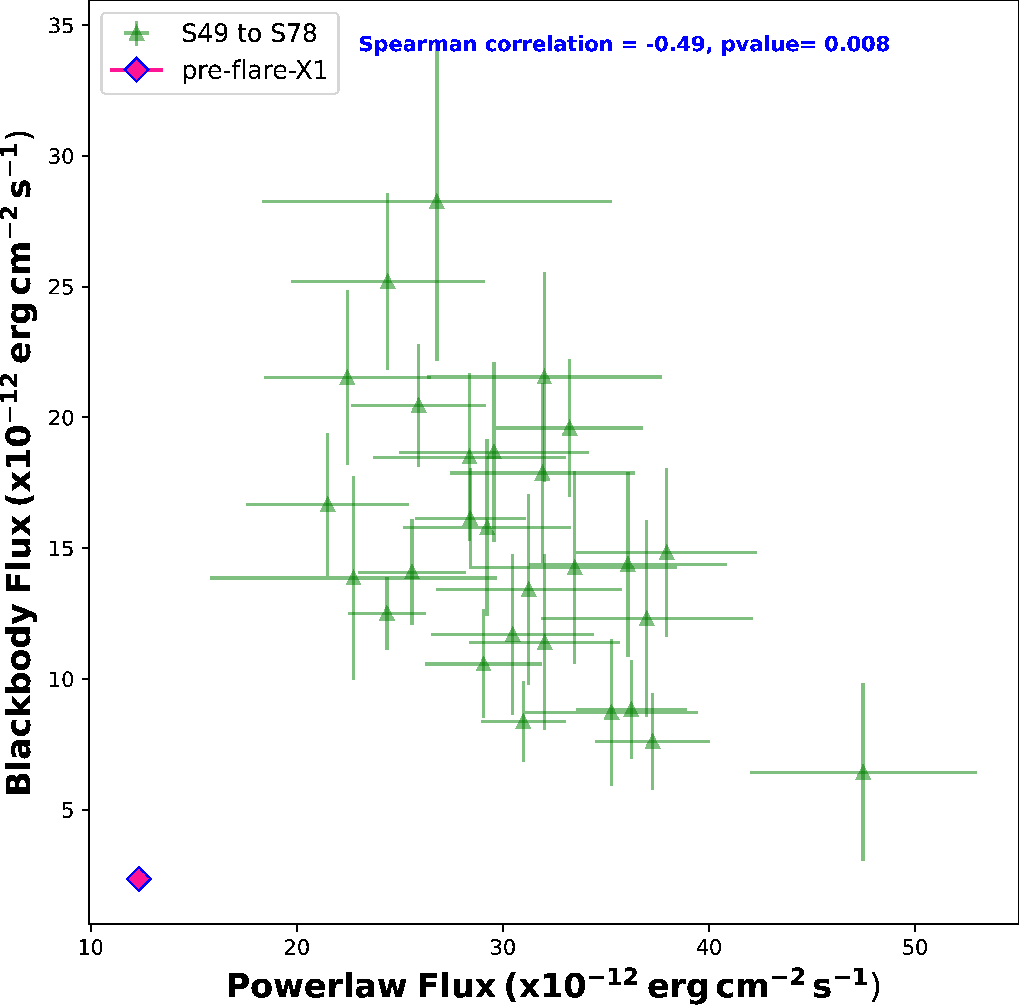}
     \includegraphics[width=5.9cm]{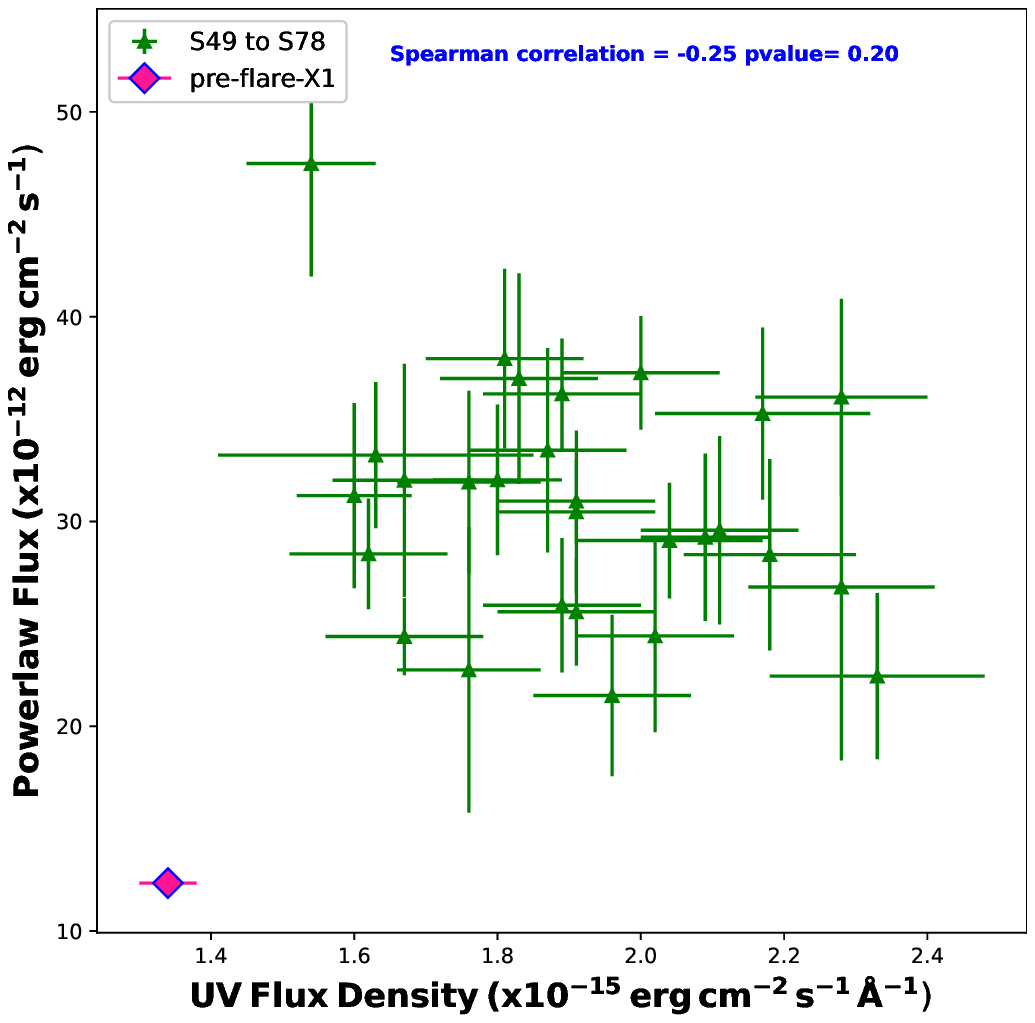}
     \includegraphics[width=5.9cm]{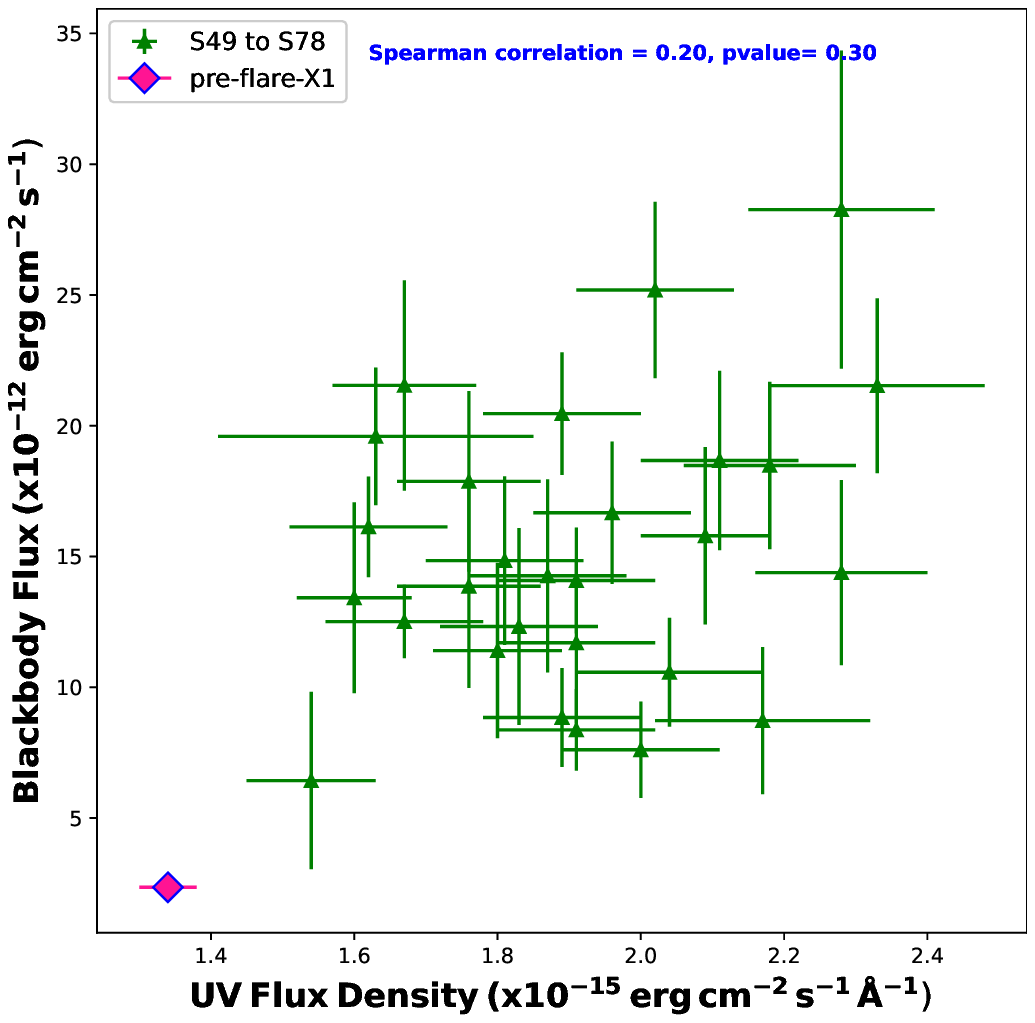}
    \caption{ {Left:} The powerlaw ($0.3-10\kev$) flux vs the blackbody ($0.3-10\kev$) flux plot during the bright soft state (i.e., observations S49-S78) exhibits no significant correlation. The bright-soft-state plotted in green, and the pre-flare XMM data plotted in magenta for comparison.
    {Middle:} The powerlaw $0.3-10\kev$ flux vs UVW2 flux density plot and {Right:} The blackbody $0.3-10\kev$ flux vs UVW2 flux density plot during the same period as in the left panel similarly shows no statistically significant correlation. }
    \label{fig:po_bb_uv_corr}
\end{figure*}


\subsection{Correlations between soft X-ray excess, the coronal and the UV flux }

We find a significant correlation between $F_{0.3-2.0\kev}$ and $F_{2-10\kev}$ in the post-flare bright-soft-state of the source. Figure~\ref{fig:soft_hard_correlation} right panel shows the correlation  between these two quantities for the observations S33-S78 when the soft X-ray flux steadily increased (green data points). In comparison, we note that there was no correlation between the soft and hard X-rays during the `initial flare' where the non-parametric Spearman rank correlation coefficient is 0.60 with a confidence level $\sim 98.0\%$(See Fig.~\ref{fig:soft_hard_correlation} left panel). To understand if there is indeed any relation between the soft excess and the power law emission (in the current bright-soft-state), we correlated the respective fluxes obtained from the best fit models, black-body and power law respectively. We note that we required a blackbody component to model the soft excess only for the observations S49 to S78. We do not find any statistically significant correlation as shown in Fig \ref{fig:po_bb_uv_corr} left panel. This indicates that the soft excess and power law variations are independent of each other, and the apparent correlation between the $0.3-2\kev$ vs $2-10\kev$ flux is driven by the power law component which is very soft ($\Gamma\sim 2.5-3$). Similarly, we find no correlations between the UV and the power-law flux and UV and blackbody flux (See Fig.~\ref{fig:po_bb_uv_corr} middle and right panel). 



We also checked for individual soft X-ray flares between different observations and found e.g., the soft X-ray has flared by a factor of a few times in the span of a few days: between S45 and S46, S48 and S49, S50 and S51, and between S65 and S66 (See Table \ref{Table:xray_obs}). However, the hard X-ray flux not always shown an increase corresponding to a soft X-ray increase e.g., between S50 and S51, the soft X-ray flux increased but the hard X-ray flux decreased. This is demonstrated in the lack of correlation between the soft excess flux and the power-law flux.


\subsection{Evolution of hardness ratio}
\begin{figure}
    \includegraphics[width=9cm]{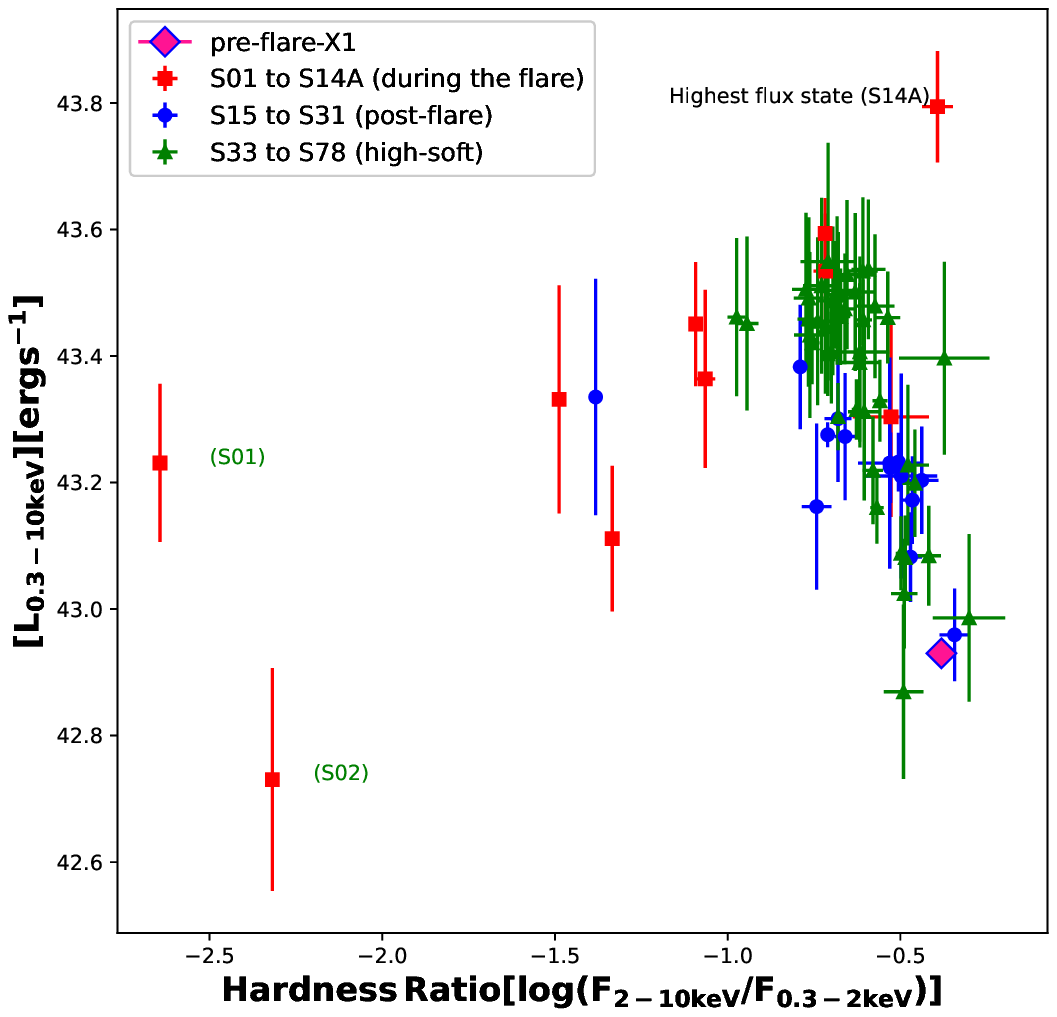}
    \caption{Hardness–intensity diagram for 1ES 1927+654. The `initial flare' S01-S14A observations are plotted as red squares. The post-flare states S15-S31 (when the source got back to the pre-flare state) are plotted as blue circles. The bright-soft-state S33-S78 are plotted as green triangles. (see Table \ref{Table:xray_obs} for details).}
    \label{fig:lhard_hr}
\end{figure}

Interestingly, the hardness ratio (HR) (defined as $F_{2-10.0\kev}/F_{0.3-2.0\kev}$) for this source always varied between $0.002\pm0.001$ (during the initial flare S01) to $0.500\pm0.044$ (during S36). If we compare these values to the pre-flare value of $0.417\pm0.021$ (X1) we see the HR 
has always been either lower or within errors compared to the pre-flare value. During the `initial flare', the HR showed both hard-when-bright (while rising) and soft-when-bright (while coming back) behaviour. The HR increased at the observation when the corona was revived in Oct 2018. In the post-flare state, during observations S15-S33, when the source showed minimal X-ray variations, the HR still varied considerably within this range. See Fig.~\ref{fig:xray_uv_alpha_ox} panel 3. We do not detect any fixed pattern for the HR variability in response to the X-ray flux changes. Nor do we find any significant correlation between $L_{\rm 0.3-10\kev}$ and  HR (See Fig. \ref{fig:lhard_hr}). During the recent bright-soft-state observed from S33, the HR varied considerably between these limits. Currently (S78), in the bright-soft-state, we find the HR to be $\sim 0.172$


\subsection{Correlation between $\Gamma$ vs $L_{2-10\kev}$}



We plotted the power law slope $\Gamma$ and the hard X-ray luminosity ($L_{2-10\kev}$) for the bright-soft-state (Fig.~\ref{fig:lhard_gamma}). It is evident from the figure that in the post-flare state, the power-law slope is insensitive to the variations in the $2-10\kev$ luminosity. 


\subsection{The evolution of $\alpha_{\rm OX}$ }

The ratio between the X-ray and UV, which we refer to as $\alpha_{\rm OX}$, is calculated from the ratio of the monochromatic fluxes in the UV ($2500$\AA) and X-rays ($2\kev$), i.e., $\alpha_{\rm OX}= -0.385 \log(F_{\rm 2\kev}/F_{\rm 2500\AA})$ \citep{tananbaum1979,lusso2010}. This is an important diagnostic parameter to understand if the accretion disk and the X-ray-emitting corona are physically connected. The $\alpha_{\rm OX}$ had attained a very high value of $\sim 2$ during the `initial flare'; however, in the post-flare state, the values are mostly consistent with that of the pre-flare state. In the recent bright-soft-state, the $\alpha_{\rm OX}$ has dropped.  This is because there is no change in the UV flux, while the X-ray flux has increased. Figure \ref{fig:alphaOX} shows the relation between $\alpha_{\rm OX}$ and $L_{\rm 2500\AA}$ and we find that the recent bright-soft-state data points denoted by blue dots deviate considerably from the standard AGN behaviour. We plot the `highest flaring state' data point on this plot to give an idea of where the source was in this phase space during the `initial flare'.



\subsection{The 5-GHz Radio observations}


The point source radio flux density at 5 GHz was at a minimum (a factor of four times lower than the pre-flare value) when the X-rays were just starting to increase. Over a period of
the next one thousand days, the radio flux density gradually increased \citep{2022ATel15382....1Y}, which coincided with the time period when the UV and X-rays returned to their pre-flare values (see Fig \ref{fig:xray_uv_alpha_ox}). 
However, the latest VLBA observation in August 2022 showed a decrease in the core radio flux density compared to March 2021. This decrease also coincides with the latest 
rise in the soft X-ray flux (bright-soft-state). The radio flux density could be related to either the hard or the soft X-ray or both in a more
complex way.

\begin{figure}
    \centering
     \includegraphics[width=8.5cm]{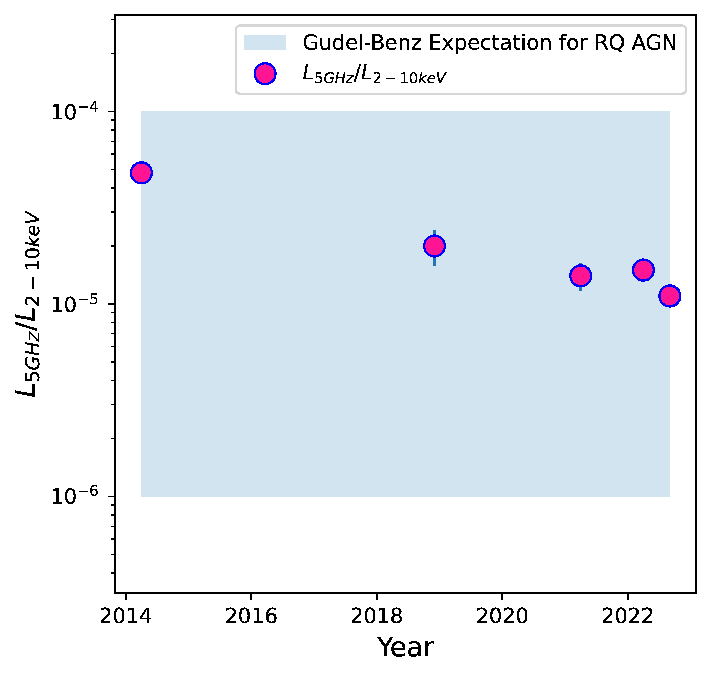}
    \caption{ The light curve of the ratio of the 5GHz mono-chromatic central radio luminosity with the $2-10\kev$ luminosity, popularly known as the G\"udel-Benz relation. See Table \ref{Table:gb} for details. The $0.3-2.0\kev$ X-ray flux vs the 5GHz monochromatic radio flux exhibits no correlation. See Table~\ref{Table:xray_obs} and Table~\ref{Table:gb} for the X-ray and Radio flux values. The shaded region represents the standard range of the G\"udel-Benz relation for radio-quiet AGNs. }
    \label{fig:gudel_benz}
\end{figure}

The ratio between the radio and the X-ray (known as the G\"udel Benz relation) has been plotted in Fig.\ref{fig:gudel_benz}. We find that the ratio constantly decreased during the
initial flare, it picked up a little in 2021 and then started decreasing again. The values of the ratio are well within the spread of the values found typically in radio-quiet AGN in the Palomar-Green sample studied by \citet{laor08}.
We see only a minor change in the morphology (size became smaller) of the extended emission but not in intensity.



\subsection{The optical spectra}

\begin{figure*}
    \includegraphics[width=17.5cm, height=8cm]{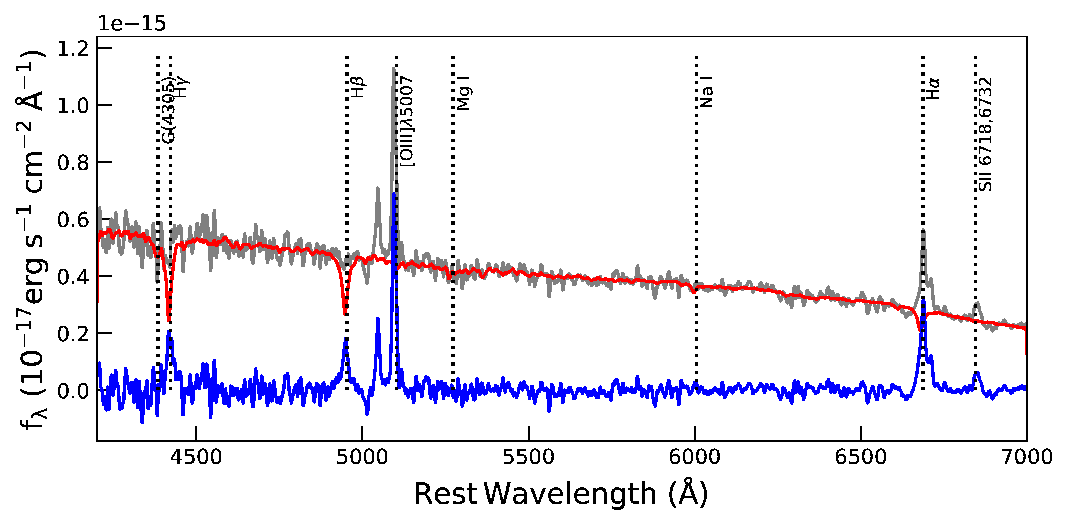}
    \caption{Optical spectrum of 1ES~1927+654 obtained with HCT. The data is shown in grey, the decomposed stellar template is shown in red, and the host subtracted spectrum is shown in blue. The spectrum was smoothened by a 5-pixel box car for visualization purposes only.}
    \label{fig:HCT_optical_spectra_first}
\end{figure*}

\begin{figure*}
    \includegraphics[width=12.5cm, height=4.5cm]{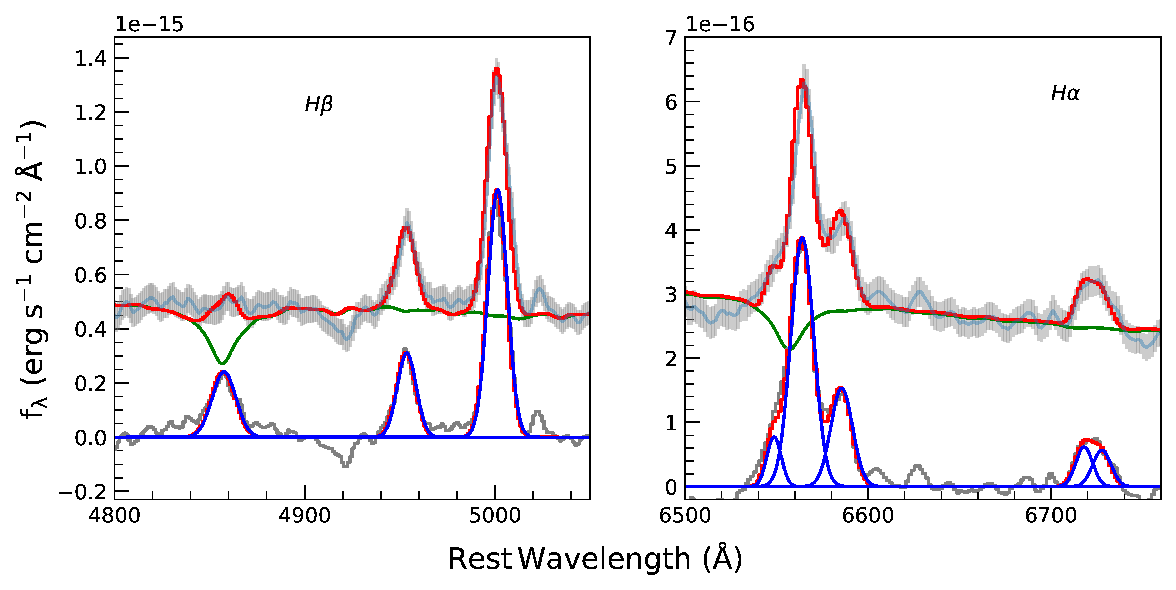}
    \includegraphics[width=5.5cm, height=4.5cm]{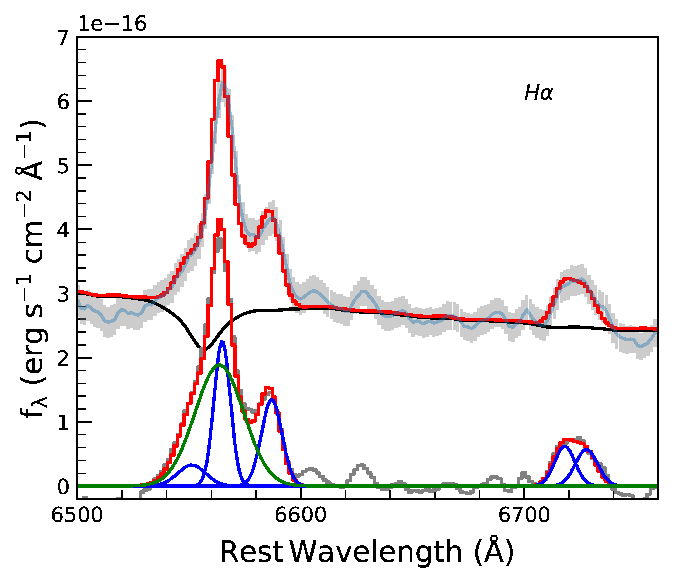}
    \caption{Emission line fitting of the optical spectrum obtained by HCT. The plot shows observed data (solid line with flux error in gray), best-fit total model (red), stellar contribution (green), and host-subtracted spectrum (gray) with individual narrow lines (Gaussian). Without any broad component in H$\beta$ and H$\alpha$ in the left and middle panels respectively and with one broad (green -Gaussian) component in H$\alpha$ in the right panel. The spectrum was smoothened by a 5-pixel box car for visualization purposes only.}
    \label{fig:HCT_optical_spectra_second}
\end{figure*}

Both the optical spectra from LDT and HCT show mostly narrow emission lines, as reported in our earlier work. We also detect broad H$\alpha$ with a typical FWHM of $1172\pm327\kms$. We do not see any change in the narrow or broad emission line intensities and
the line ratios from our earlier work \citep{laha2022} except for OIII where the line intensity has nearly doubled. The reduced spectra and the best-fit model from HCT are shown in Figure \ref{fig:HCT_optical_spectra_first} and Figure \ref{fig:HCT_optical_spectra_second} respectively showing line intensities consistent with \cite{laha2022}.

\begin{figure}
    \includegraphics[width=8.5cm]{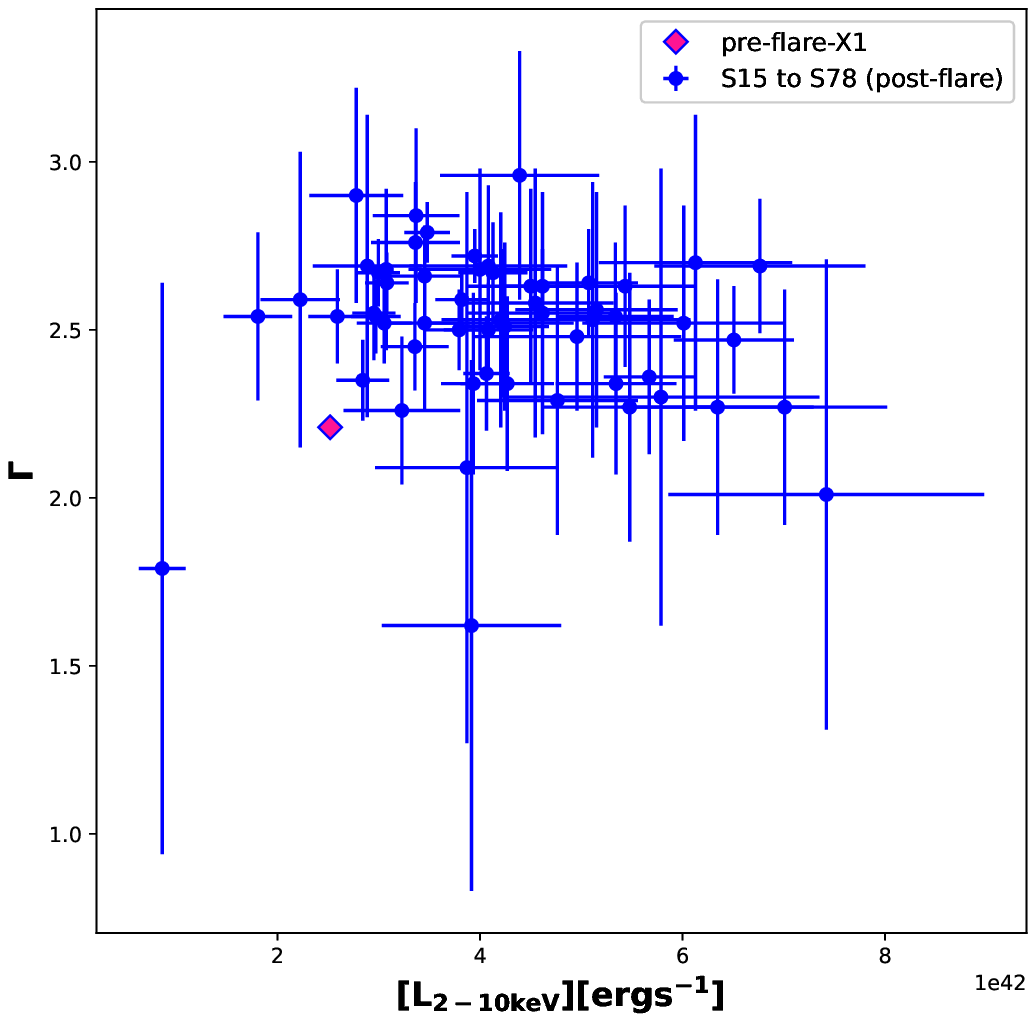}
    \caption{The photon index $\Gamma$ and the $2-10\kev$ X-ray luminosity plot during the bright-soft-state that is S33-S78. 
    The $\Gamma$ is relatively insensitive to the hard X-ray luminosity. 
    }\label{fig:lhard_gamma}
\end{figure}


\begin{figure*}
   \centering
\hbox{
\includegraphics[width=9cm]{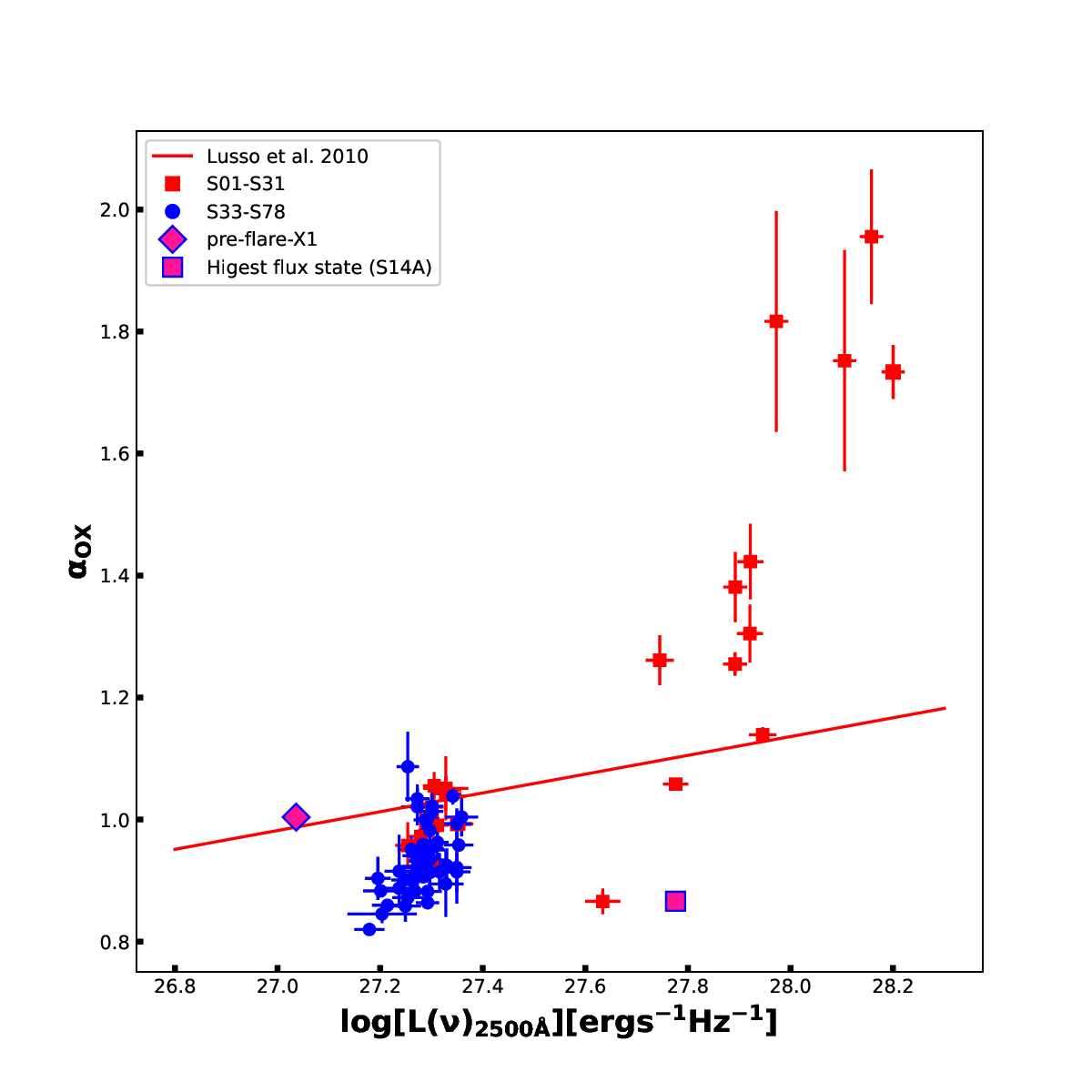}
\includegraphics[width=9cm]{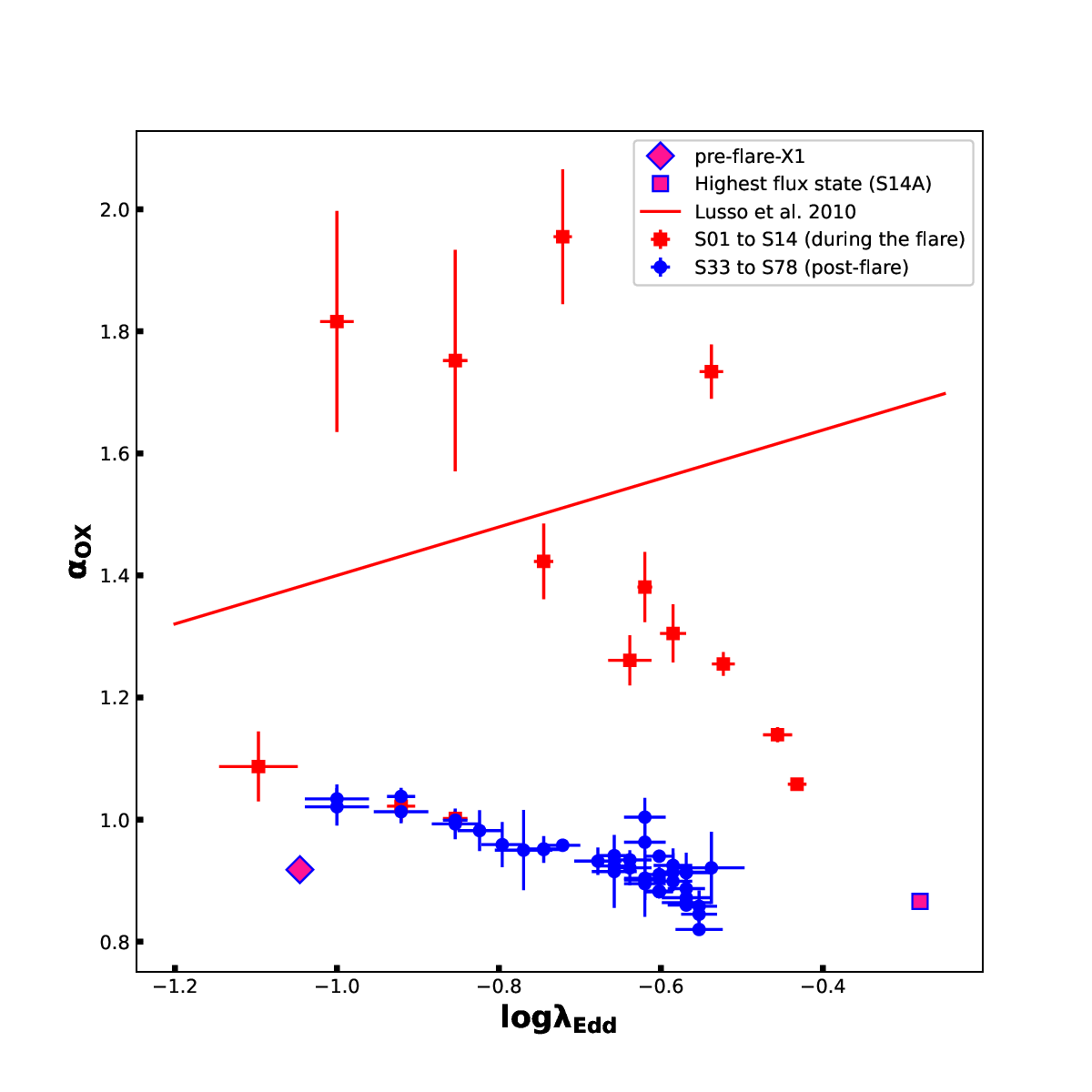}}
    \caption{ {\it Left:} The $\alpha_{\rm OX}$ vs $L_{2500\angs}$ for the different \swift{} observations of the source 1ES~1927+654 as reported in Table \ref{Table:xray_obs} and \cite{laha2022}. The red line is the best-fit correlation from \cite{lusso2016} representing the standard AGN disk-corona relation. The pink diamond represents the 2011 pre-CL state of 1ES~1927+654. {\it Right:} The evolution of $\log(\lambdaedd)$ and the $\alpha_{OX}$ during the initial flare (red squares) and the post-flare state (blue circles). See Table~\ref{Table:lambdaedd} for details). The solid red line is the generic AGN behaviour, as obtained by \citep{lusso2016} for a sample of AGNs. We do not see any correlation between these parameters either during the initial flare or the post-flare states. 
     }
    
    \label{fig:alphaOX}
    
\end{figure*}

\section{Discussion}\label{sec:discussion}

We have followed up the \cl{} 1ES~1927+654 with X-ray and UV observations with \swift{}, radio observation from VLBA, and optical observation from LDT and HCT. Our earlier paper \cite{laha2022} reported the observations till 31st December 2021, encompassing the entire phase of the `initial flare'. Here we report the multi-wavelength observations from 1st January 2022 till 5th May 2023, particularly highlighting the recent bright-soft-state of the source.  In light of our multi-wavelength observations, we address the following scientific questions.


\subsection{The recent bright-soft-state and the origin of soft X-ray excess}\label{subsec:soft-excess}

The soft X-ray excess (SE) is the excess emission (over the power law) in the soft band $0.3-2\kev$ in an AGN spectrum which typically takes the shape of a black body with temperatures ranging $0.1-0.3\kev$. The origin of the SE is still highly debated, \citep{noda_mrk1018,garcia_mrk509,ghosh_eso141,ghosh_es0511} as different sources show different spectral and timing behaviour of the SE with respect to the other bands of the AGN continuum such as the UV (accretion disk) and the $2-10\kev$ (power law). The two most popular models describing the SE are (a) Thermal Comptonization from warm ($kT_{e}\sim 0.1-0.5\kev$) optically-thick ($\tau\sim 10-20$) corona \citep{mehdipour2011,Done2012,ghosh_1h0323,Petrucci2018,Petrucci2020} and (b) reflection of the primary continuum from ionized disk \citep{Garcia2014,ghosh_pks0558, garcia2019}.

The thermal Comptonization assumes the existence of a warm ($kT\sim 0.1\kev$) optically thick medium above the accretion disk. A fraction of the accretion disk photons interact with this plasma and gets up-scattered to create the black body shaped soft excess. Therefore this model points towards a direct link between the UV and SE flux. This model predicts an increase in the UV should be related to an increase in the soft X-rays and vice-versa.

The relativistic reflection, on the other hand, assumes that the SE arises out of the reflection of the hard X-ray photons from the primary continuum emission from the ionized accretion disk. The individual emission lines arising out of the reflection get smoothed out (broadened beyond detection) by the gravitational effects. This model, therefore, assumes a direct link between the $2-10\kev$ flux and the SE flux.

The SE behaviour in source 1ES1927+654 is very unique. During the `initial flare' (during 2017-2019, reported in \citealt{Ricci2021,laha2022,masterson2022}), the source started off very soft when the UV was gradually decreasing. When the $2-10\kev$ band X-rays completely vanished, there was still some traces of the SE as detected by \xmm{}  \citep{Ricci2021}. This phase of the SE was characterized by an unusually varying lower temperature (kT$\sim 0.08-0.1\kev$) black body. The temperature  was lower than that usually measured for this source in the pre-flare state (kT$\sim 0.2\kev$). The presence of SE in the absence of a power law component is inconsistent with the reflection scenario because, without the primary power law, there should not be any reflection in the first place. This is also unusual in the warm Comptonization scenario because the UV flux was very high at the time when the SE was the lowest unless the warm Comptonizing corona also vanished in some way. When the X-rays revived with a flare in Oct 2018 (see Fig. \ref{fig:xray_uv_alpha_ox}) the UV was still monotonically decreasing with a power law decline $y\propto t^{-0.91}$. There was no correlation between the UV and X-ray fluxes, or even between the soft and the hard X-ray fluxes (see Fig. \ref{fig:soft_hard_correlation} left panel). 
These results make it hard to describe the SE of this source with either of the two models mentioned above.

\begin{figure*}
    \centering
     \includegraphics[width=8.5cm]{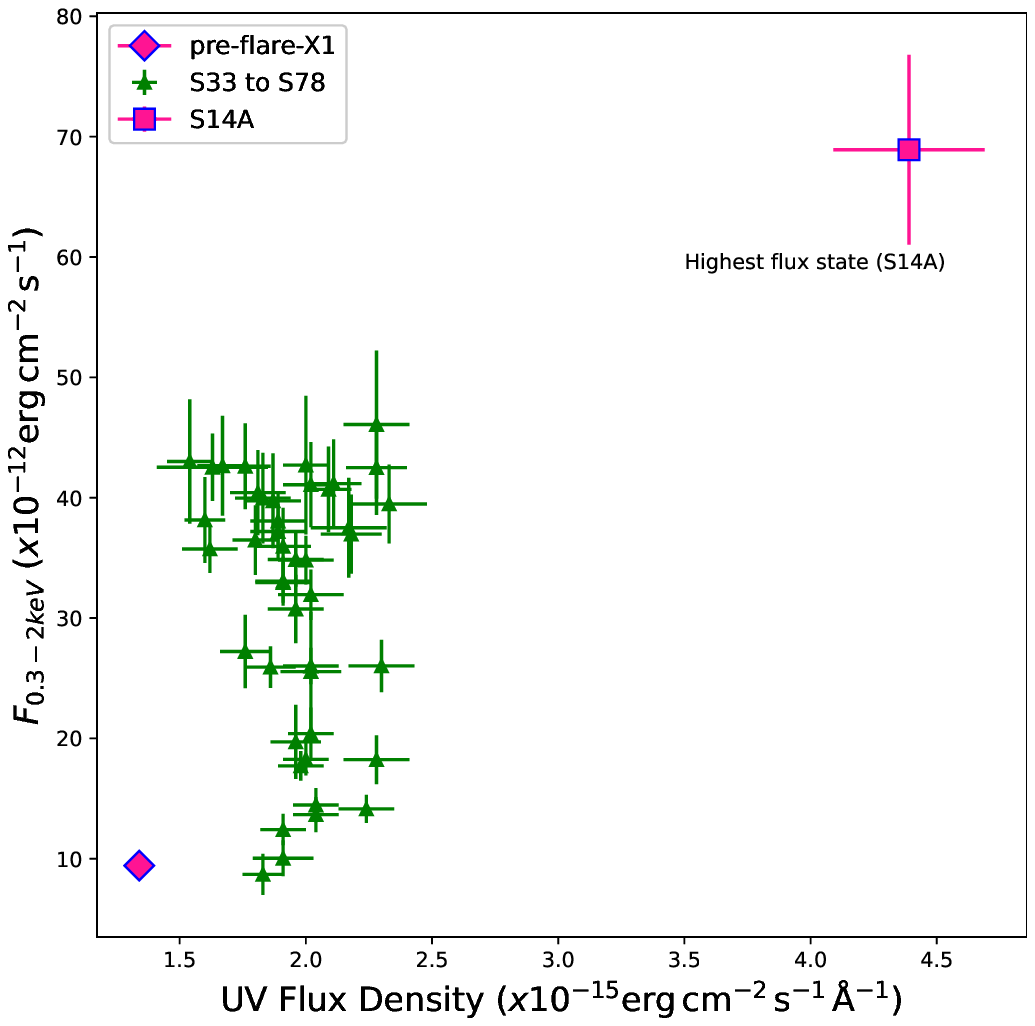}
     \includegraphics[width=8.5cm]{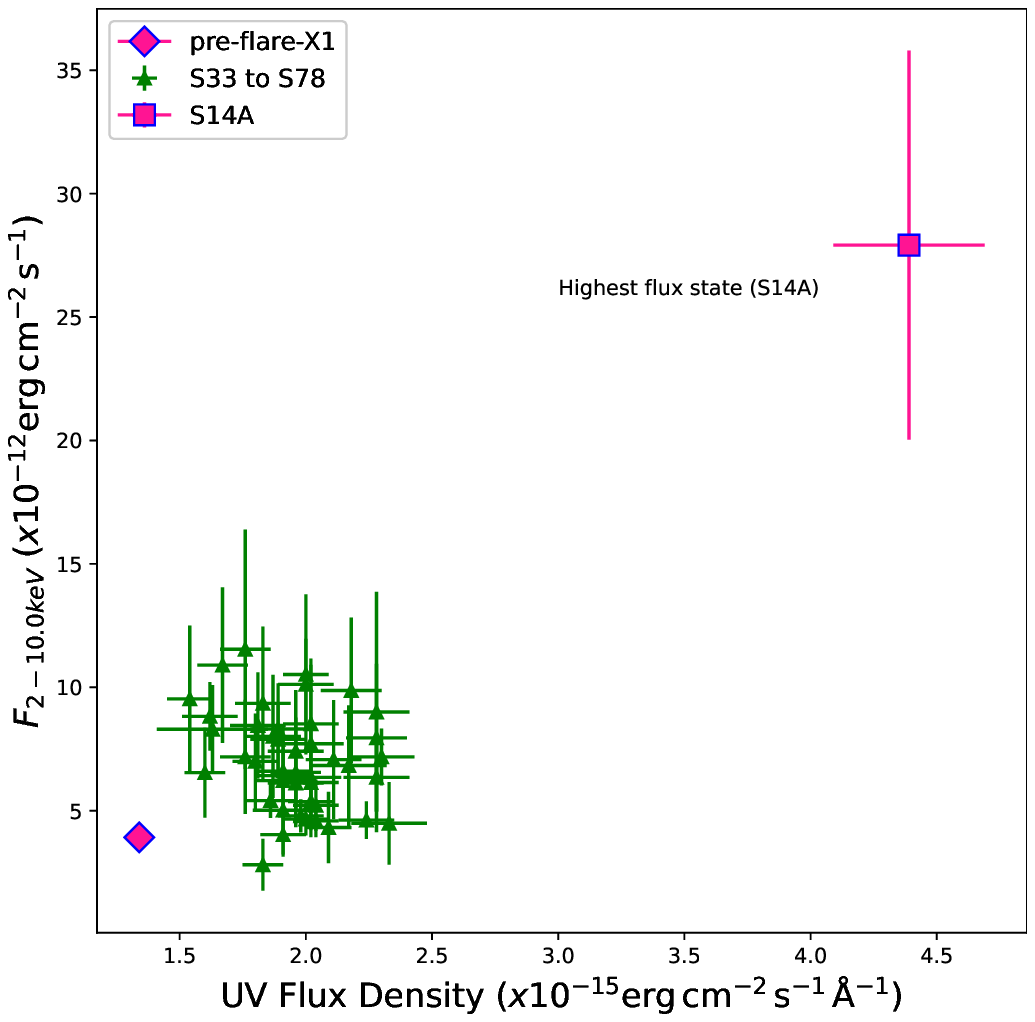}
    \caption{ {Left:} The soft X-ray $0.3-2\kev$ flux vs the UVW2 flux density during the bright soft state (i.e., observations S33-S78) exhibits no significant correlation. The bright-soft-state plotted in green, the pre-flare XMM data plotted in magenta and the highest flux state in magenta.See Table \ref{Table:xray_obs} for details. {Right:} The hard X-ray $2-10\kev$ flux vs UVW2 flux density during the same period as in the left panel similarly shows no statistically significant correlation. }
    \label{fig:uv_vs_soft_hard}
\end{figure*}

During the recent bright-soft-state since May 2022, which became more prominent in Dec 2022, we find that the UVW2 flux density changes are confined within a fluctuation range of $\le 30\%$. The mean value of the UV flux in the bright-soft-state is $1.96\times 10^{-15}$ and a standard deviation of $0.19\times 10^{-15}$ (in units of $\funita$). While the UV flux shows minimal change, the soft X-ray flux has increased by a factor of five compared to the pre-flare state. The hard X-ray flux has also increased by a factor of $\sim 2$ but has shown substantial fluctuations in the latest observations (see Fig. \ref{fig:xray_uv_alpha_ox_zoomed} for details). We do not see any correlations between the UV and soft X-rays and UV and hard X-rays (see Fig. \ref{fig:uv_vs_soft_hard}). This is in agreement with the lack of correlation observed between the soft excess flux (modelled by the blackbody) and the coronal flux (modelled by the power-law) in the recent bright-soft-state (Fig.\ref{fig:po_bb_uv_corr}). These results are inconsistent with both the models describing the soft excess as the reflection from the ionized disk and/or thermal Comptonization scenario. 

Previous studies using broadband spectroscopy with \xmm{} concluded that the $1\kev$ emission feature (detected in \xmm{} and \nicer{} observations) could arise out of the reflection of the hard X-ray photons of the accretion disk \citep{masterson2022}.  They suggested relativistic reflection happening during the TDE, resulting in the $1\kev$ emission line. The spectroscopic fits could not distinguish between the thermal Comptonization and the reflection models for the origin of the SE. 

Other CL-AGN sources have also shown the absence of correlation between the X-rays and the UV and between the soft and the hard X-rays. For example, Mrk~590 \citep{Ghosh_mrk590} where we found that the SE completely vanished when the power law and the UV were still dominant. We also note that in 1ES1927+654 since the start of the bright-soft-state, the $\Gamma$ has not varied significantly, staying between $2.01^{+0.50}_{-0.90}$ and $2.96^{+0.36}_{-0.37}$, consistent within the large errors.

Now the question is: what is pumping the energy to the X-rays if the UV flux, and hence the standard accretion rate, is so silent? To understand this, we need to have some estimate of the energy. First, we calculate the energetics of this system in one of the shortest variability time scales, which in this case is 4 days (cadence of our {\it Swift} observation). We note that the highest soft and hard X-ray flux change happened between S48 and S49 (gap of 4 days). Assuming a constant increase profile, we estimated the soft band energy pumped within four days is equal to $9.7\times 10^{47} \rm erg$. During that time, the total hard X-ray energy pumped is equal to $3.7\times10^{47} \rm erg$. While the energy pumped in UV amounts to $1.9\times10^{47} \rm erg$, which roughly equals $14\%$ of the total X-ray energy pumped during this period.

\begin{figure*}
   \centering
\hbox{
\includegraphics[width=9cm]{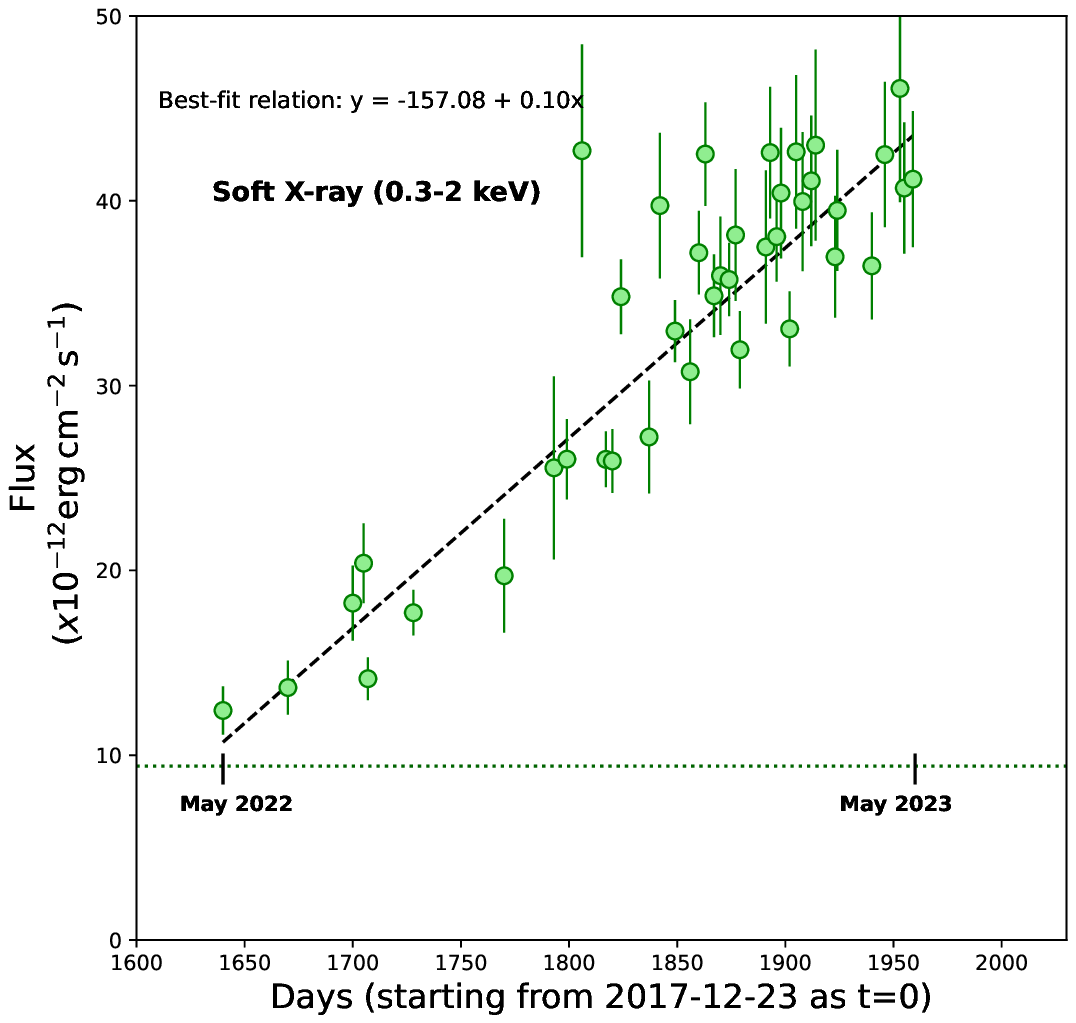}
\includegraphics[width=9cm]{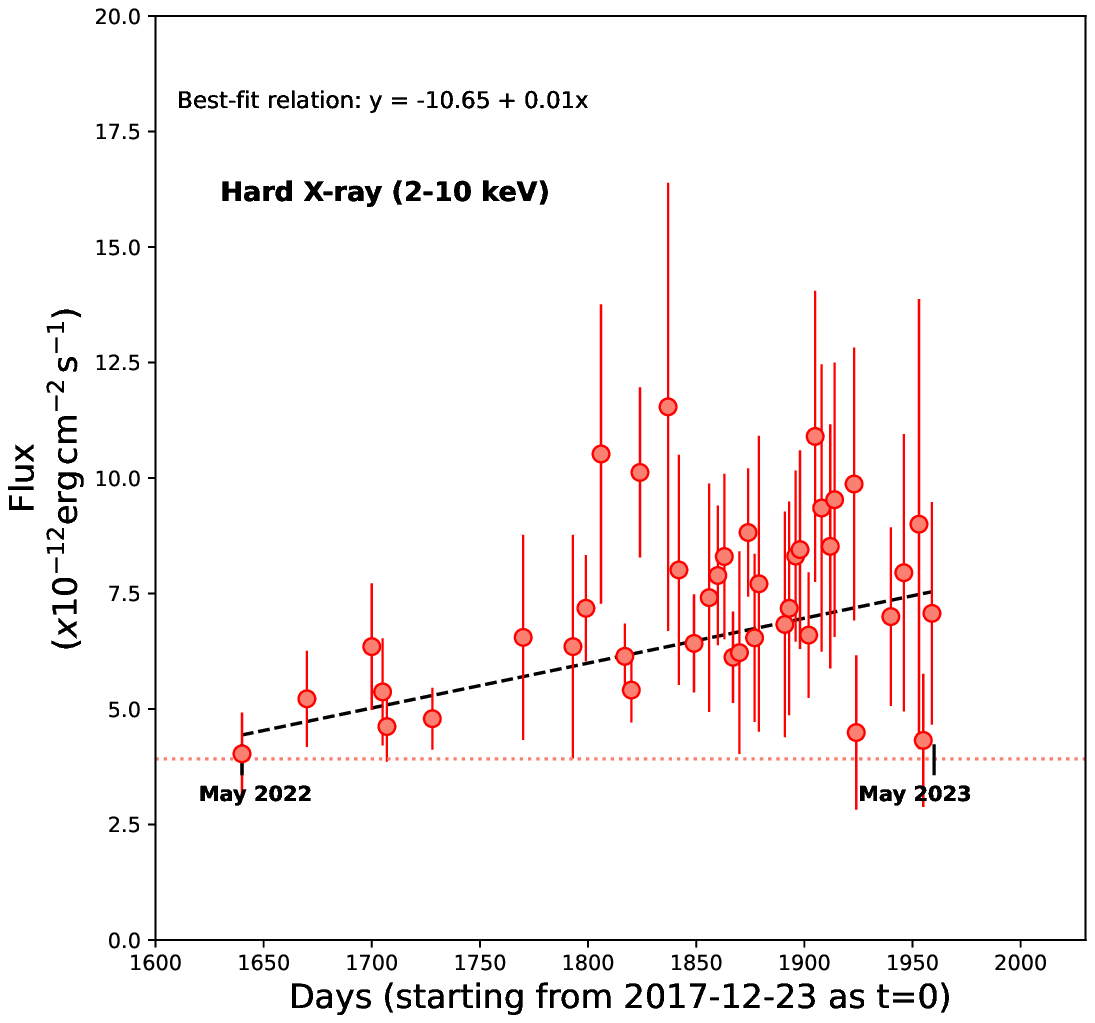}}
    \caption{ {\it Left:} The linear regression fit to the soft X-ray energy increase during the recent bright-soft-state. {\it Right:} Same as left, but for the hard X-rays. We have estimated the total integrated energy input in the soft and hard X-rays during this time period by integrating under the best-fit line, as shown above. See Section \ref{subsec:soft-excess} for details. The dotted horizontal line represents the pre-flare flux values obtained from the 2011 \xmm{} observation.
     }
    
    \label{fig:integrated_energy}
    
\end{figure*}

Now let's consider the energy pumped in a longer time scale since the rise of the bright-soft-state phase, which is between S35(May 2022) and S76(May 2023). During these eleven months (344 days), the energy pumped in the soft and the hard bands are $3.57\times 10^{50} \rm erg$ and $0.59\times 10^{50} \rm erg$, respectively. During this one year long time period the UV band did not show any significant change at all (see Fig. \ref{fig:xray_uv_alpha_ox_zoomed}), with fluctuations consistent within one standard deviation around the mean value. Fig \ref{fig:integrated_energy} shows the soft and the hard X-ray band rise and the best fit linear regression line. The integrated energy for soft and hard bands quoted above (for 11 months) has been estimated by the area under the best fit linear regression curve. 

In addition, we can estimate the mass accretion rate required to create the luminosity in the soft X-rays from the standard accretion theory, $L=\eta \dot{M}c^2$, where $L$ is the accretion luminosity, $\eta$ is the accretion efficiency, $\dot{M}$ is the mass accretion rate, and $c$ is the speed of light. This is assuming that whatever is creating the soft X-rays is accreting and deriving its energy from the gravitational binding energy. If we assume a value of $\eta=0.1$ for standard AGNs, we get an accretion rate of $\sim 0.003 \msolyi$ needed to create this bright-soft-sate.

From the energetics, it is evident that whatever is producing the SE is pumping out more energy than either the UV or hard X-ray source. Since the energy source presumably is ultimately accretion of matter onto the SMBH, the SE emitting region must be receiving the majority of this energy.
This could mean that the SE region is either the dominant accreting flow or there is a conduit pumping the accretion energy from the disk to the SE region rather than radiating it inside the disk.

If warm Comptonization produced the soft excess, then one would expect a direct relation between the UV luminosity and the soft X-ray luminosity e.g., as observed in NGC~3516 \citep{mehdipour2022_NGC3516}, but we do not find in 1ES~1927+654. Moreover, one would expect to observe absorption in the soft X-rays due to the atomic opacity in the warm Comptonizing corona \citep{garcia_mrk509}, which we do not see in the {\xmm} EPIC-pn data of the source in the bright-soft-state (Ghosh et al. in preparation). The reflection scenario, on the other hand, cannot explain a weaker ($\sim 2$ times) hard X-ray variation (which is the primary flux) compared to the stronger ($\sim 5$ times) consistent rise of soft X-ray, with the soft X-rays gaining $\sim 10$ times more energy than the primary power law in a matter of 11 months. This is also true if we see the initial flare \citep{laha2022} where no correlation between the soft and hard X-ray was found. 

Therefore, in this source, we need a different mechanism to produce the soft X-ray excess. We conjecture that the soft excess in this source may not be the canonical soft excess we find in typical AGN. This is supported by several observational results from both the initial flare (Dec 2017- Dec 2019) and the bright soft state (May 2022- May 2023), such as (1) no correlation between the SE and other X-ray and UV components (2) SE modeled by an unusually varying lower temperature black body (kT$\sim 0.08-0.1\kev$) in initial-flare when the coronal emission vanished, (3) The current consistent rise in the SE having a black body temperature mostly constant but on the higher side (kT$\sim 0.15-0.2\kev$).

\subsection{The relation between $\alpha_{\rm OX}$ vs $L_{\rm 2500\AA}$ and $\alpha_{\rm OX}$ vs $\lambdaedd$}

The relation between $\alpha_{\rm OX}$ and $L_{\rm 2500\AA}$ has been found to be tightly related for AGN across cosmic timescales \citep{lusso2010,lusso2016}, which implies a close relation between the accretion disk (UV) and the corona (X-ray). Figure \ref{fig:alphaOX} left panel shows the data points for the source 1ES~1927+654 during the initial flare (in red) and the bright-soft-state (in blue), and in neither of the cases, the source behaviour follows the commonly detected correlation (red line). We note that the pre-flare state (in pink diamond) was close to that of the expected slope, and possibly the AGN disk corona emission was related at that time. We also note that when the source went back to the pre-flare state after the initial flare, the red dots and blue circles were near the expected slope. However, in neither the `initial flare' nor the bright-soft-state, the disk corona relation holds.

Similarly, from Fig. \ref{fig:alphaOX} right panel, we note that the $\alpha_{\rm OX}$ vs $\lambdaedd$ show no fixed pattern. \cite{ruan2019} compared the observed correlations between $\alpha_{\rm OX}$ and $\lambdaedd$ in AGN (including low accretion states $\lambdaedd\sim 10^{-2}$) to those predicted from the observations of X-ray binary outbursts. They found that the observed correlations in AGN are very similar to the accretion state transitions in typical X-ray binary outbursts, including the inversion of this correlation at $\lambdaedd\sim 10^{-2}$. The right panel of Fig. \ref{fig:alphaOX} shows that 1ES~1927+654 never went below $\log\lambdaedd<-1.2$, and hence we are confined to only the high accretion phase space of the study by \cite{ruan2019}. The bright-soft state is visible (red squares at the top), but as the UV faded, the X-rays grew in strength, leading to the `high-hard' state (lowest pink square at the bottom right). 
The red line represents the general AGN behaviour as obtained by \cite{lusso2016}.



\subsection{$\Gamma$ vs $\lambdaedd$ and $\Gamma$ vs $L_{\rm 2-10\kev}$ evolution}

In a large sample of $\sim 7500$ AGN \cite{sobolewska2009} found that the average spectral slope $\Gamma$ does not correlate with source luminosity or BH mass while it correlates positively with average $\lambdaedd$. From Fig. \ref{fig:gamma_vs_eddington}, we find that in the `initial flare', there was no correlation between $\Gamma$ and $\lambdaedd$ (red data points). However, the bright-soft-state is different (in blue points). We find that although there has been an increase in the $\lambdaedd$ in this phase, the power law $\Gamma$ is very narrowly distributed around $\sim 2.5$. Similarly the $\Gamma$ vs $L_{2-10\kev}$ plot (Fig.~\ref{fig:lhard_gamma}) shows that the power-law slope is insensitive to the variation in hard X-ray luminosity. 
The conclusion is that the power-law slope is insensitive to the luminosity and $\lambdaedd$ in the bright-soft-state.






\subsection{The evolution of core and the extended radio flux density}


The physical mechanism of the origin of the unresolved core radio emission from radio-quiet AGN is still unknown \citep[see][and references therein]{2019NatAs...3..387P, 2020ApJ...904..200Y}. The possible candidates include (a) the AGN corona \citep[e.g.][]{2020ApJ...904..200Y}, (b) low-power jets \citep[e.g.][]{2021MNRAS.508.1305Y} (c) shocks from winds \citep[e.g.][]{2014MNRAS.442..784Z} and (d) rapid star formation \citep{panessa2019,panessa22,2022ApJ...938...87K}. 1ES~1927+654 has given us a unique glimpse at the core radio emission during and after the violent changing look event. From \cite{laha2022}, we note that the core radio emission was at its lowest when the hard X-ray flux was low. Gradually with time over the next 3 years \citep[till March 2022, see also][]{2022ATel15382....1Y}, we found that the radio flux density increased but never reached its pre-flare value. In a more recent observation in Aug 2022, the core radio flux density decreased, coincidentally when the soft X-ray flux started to rise (bright-soft-state). It is, therefore, clear that the core radio flux density is variable at a time scale of months and hence cannot arise out of star formation. It is likely that the core radio emission is related to the corona or the shocks from the winds generated by the violent events, a nascent evolving jet or a low-power jet decreasing in power. Future coordinated X-ray, radio and optical-UV monitoring of the source will help us resolve the nature of the radio emission.



The ratio of unresolved core radio emission at 5 GHz of radio-quiet AGN to the $L_{\rm 2-10\kev}$ luminosity follows a unique relation known as the G\"udel-Benz relation \citep[GB, see, e.g.,][]{laor08} commonly found in the coronally active stars.  Table \ref{Table:gb} lists the GB relation of the source, and Fig. \ref{fig:gudel_benz} shows the GB light curve. We find that the GB constantly reduced during the violent phase, picked up, and then again started to reduce during the bright-soft-state but still within the expected range for radio-quiet AGN. Further follow-up radio monitoring is currently being undertaken for a better understanding of the radio emission of the source.



\subsection{The optical spectra}

We do not see any significant variations in the emission line intensity in the  optical spectra between the 2021 and 2022 observations except for the OIII emission lines which have nearly doubled their intensity. We think that the light front from the December 2017 violent event may have just reached the narrow-line region. We need further optical monitoring to confirm this scenario.



\section{Conclusions}\label{sec:conclusions}

We followed up the enigmatic changing look AGN 1ES~1927+654 with multi-wavelength observations spanning a period over 1 year (Jan 2022- May 2023). Below we list the most important conclusions.

\begin{itemize}


\item We have observed a recent brightening of the soft X-ray flux since May 2022, although there is no appreciable change in the UV flux. The total energy pumped into the soft X-rays  over a period of $11$ months (20th May 2022- 5th May 2023) is $3.6\times  10^{50}$ erg, and that of the hard X-rays is $ 5.9\times 10^{49}\erg$, an order of magnitude lower. Both the warm Comptonization and the disk reflection scenarios may not be adequate to describe the soft X-rays of this source. The energetics suggest that whatever is producing the SE is pumping out more energy than either the UV or hard X-ray source. 
This implies that the SE region is either the dominant accreting flow or there is a conduit pumping the accretion energy from the disk to the SE region rather than radiating it inside the disk.

\item In the bright-soft-state we do not detect any correlation between soft X-rays vs UV and the hard X-rays vs UV. In addition, we do not see any correlation between the (a) soft excess vs the power-law, (b) soft excess vs UV, and (c) power-law vs UV. The apparent correlation between the soft and hard X-rays is driven by the power-law alone due to the soft nature of the coronal emission ($\Gamma\sim 2.5-3$).

\item The core radio emission ($<1pc$) at $5$ GHz showed an increase till March 2022 and then a dip in Aug 2022. 
It is probably linked to the coronal emission or the shocks from the winds generated by the violent events during the initial flare, a nascent evolving jet, or a low-power jet decreasing in power.

\item The G\"udel-Benz relation continuously decreased from the pre-flare state and is still within the range exhibited by most radio quiet AGN.

\item We do not detect any correlations between (a) $\Gamma$ vs $\lambdaedd$ and  (b) Hardness ratio vs $L_{2-10\kev}$, neither during the `initial flare' nor during the bright-soft-state.The power-law slope is insensitive to the luminosity and $\lambdaedd$ in the bright-soft-state.
 
 \item The data do not follow the $\alpha_{\rm OX}$ vs $L_{2500\AA}$ and the $\alpha_{\rm OX}$ vs $\lambdaedd$ relations usually observed in typical AGN where the disc-corona synergy is in place.

\item In the optical band we found that the line intensity of OIII has nearly doubled since 2021. We think that the light front from the violent event in Dec 2017 may have just reached the narrow-line-region. We need further monitoring of the source to confirm this scenario.

\end{itemize}

\clearpage{}
\section{Acknowledgements}
We thank the anonymous referee for the insightful comments that helped to better the manuscript.
RG and SL thank Missagh Mehdipour for the helpful discussion.
The material is based upon work supported by NASA under award number 80GSFC21M0002. We acknowledge the use of public data from the Swift data archive and the \swift{} team for approving the \swift{} DDT request.
MN is supported by the European Research Council (ERC) under the European Union’s Horizon 2020 research and innovation programme (grant agreement No.~948381) and by funding from the UK Space Agency. XLY is supported by the National Science Foundation of China (Grant No. 12103076) and the Shanghai Sailing Program (Grant No. 21YF1455300). K\'EG was supported by the Hungarian National Research, Development and Innovation Office (NKFIH), grant number OTKA K134213.

\section{Data availability:}

This research has used new and archival data of \swift{} observatory through the High Energy Astrophysics Science Archive Research Center Online Service, provided by the NASA Goddard Space Flight Center. This work has made use of data from VLBA (available in the NRAO Data Archive: \url{https://data.nrao.edu/}) and EVN observations (available in the EVN Data Archive at JIVE: \url{http://archive.jive.nl/scripts/portal.php}). The National Radio Astronomy Observatory is a facility of the National Science Foundation operated under cooperative agreement by Associated Universities, Inc. The EVN is a joint facility of independent European, African, Asian, and North American radio astronomy institutes.

\bibliographystyle{aasjournal}
\bibliography{main}


\appendix{}

The evolution of the X-ray spectra and the corresponding best-fit model are shown below to characterise the evolution of 1ES~1927+654 as observed by \swift{}-XRT.


\begin{figure*}
    \includegraphics[width=12cm,angle=-90]{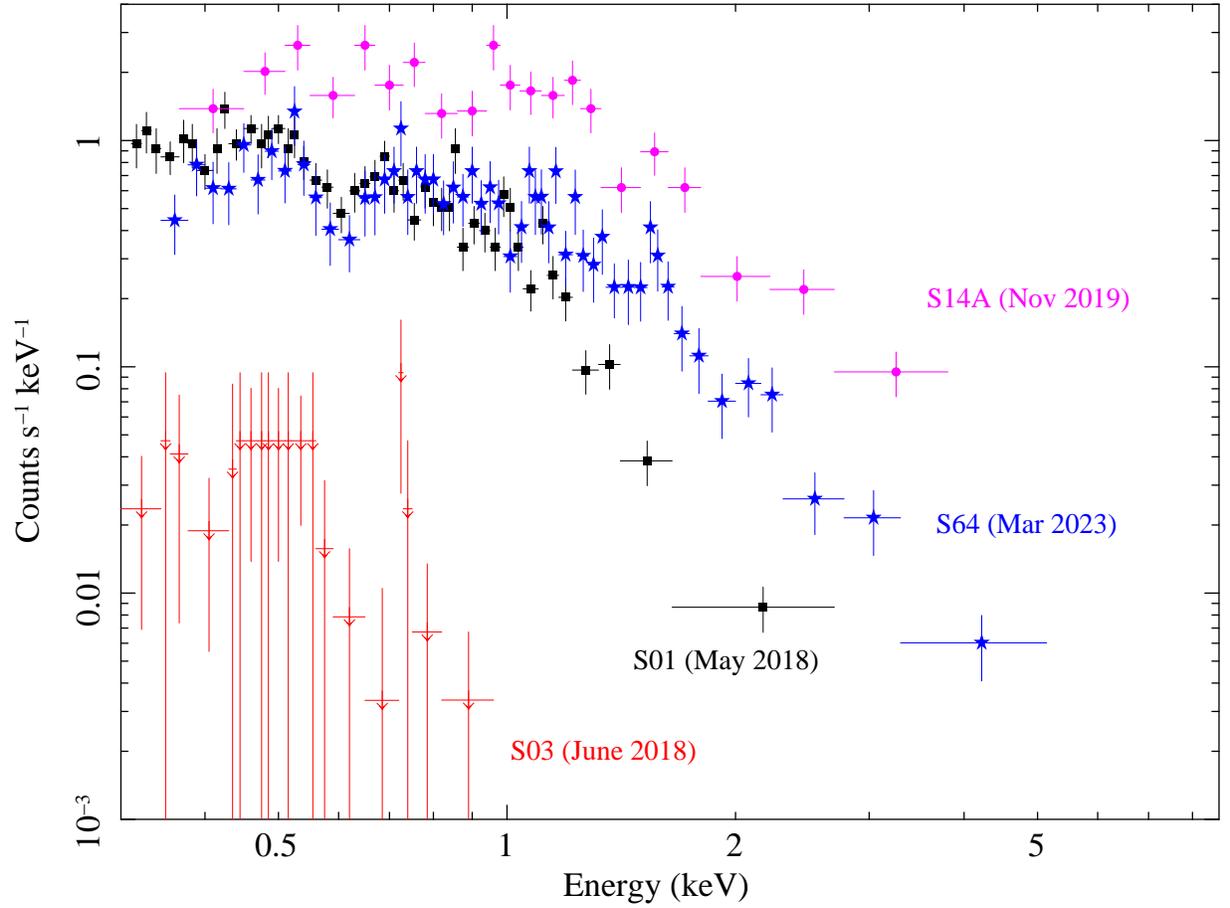}
    
    \caption{ X-ray spectral evolution of 1ES 1927+654 as observed by \swift{}-XRT. The X-ray spectrum in May 2018 is shown by black diamond points when the spectrum was very soft and the spectrum $>1\kev$ was about to vanish (corona vanished). The lowest flux state (still detectable by XRT) was captured in June 2018 shown in red triangular points. The highest flux state (S14A, see \citealt{laha2022}) in November 2019 is depicted as pink circular points. 
    The soft X-ray flux rise in Mar 2023 (during bright-soft-state) is shown in blue star points. See Table~\ref{Table:xray_obs} and \cite{laha2022} for details. }\label{fig:simultaneous_data}
\end{figure*}

\begin{figure*}
    \includegraphics[width=12cm,angle=-90]{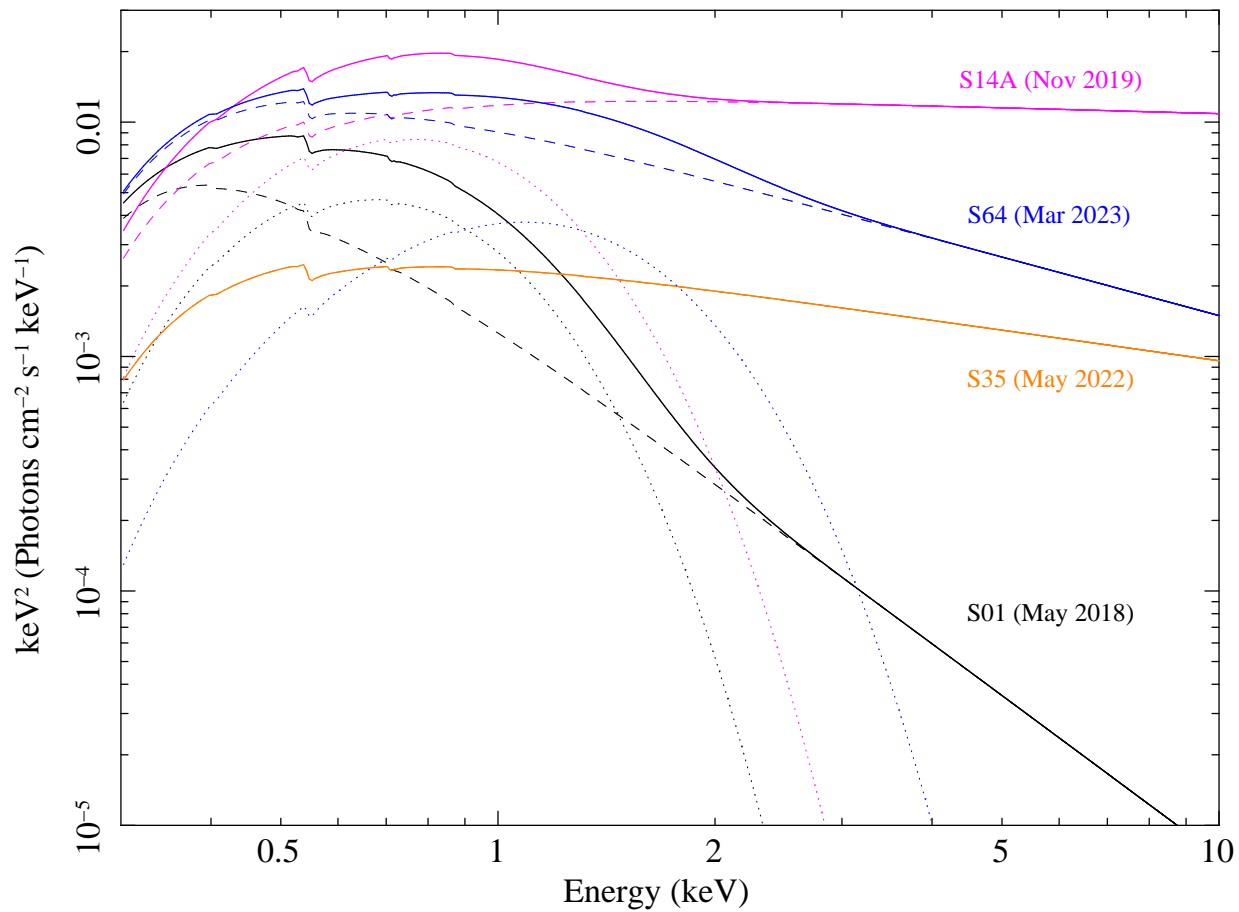}
    
    \caption{The best-fit model obtained from the spectral fitting of the \swift{} X-ray observations showing the evolution of 1ES 1927+654. We just included the best-fit models to better visualize the spectral evolution. The dotted dashed, and the solid line represent the black body, power-law, and the total spectrum respectively for each observation. The model for May 2018 is shown by a black curve featuring a soft spectrum. The highest flux state (S14A, see \citealt{laha2022}) in November 2019 is depicted as a magenta curve. S35 represents the recent observation in May 2022, where the soft and hard X-ray flux returned to the pre-flare state (as in 2011) and is shown in the orange curve. Here, we did not require any blackbody component to model the spectrum.  The soft X-ray flux rise in Mar 2023 (during bright-soft-state, S64) is shown in blue curve. See Table~\ref{Table:xray_obs} and \cite{laha2022} for details. }\label{fig:simultaneous_model}
\end{figure*}

\end{document}